\documentclass{vldb}
\usepackage{graphicx}
\usepackage{balance}  % for  \balance command ON LAST PAGE  (only there!)
\usepackage[usenames,dvipsnames]{color} % Required for custom colors
\usepackage[T1]{fontenc}    % use 8-bit T1 fonts
\usepackage{mathtools}
\usepackage{color, times}
\usepackage{xspace}
\usepackage{xcolor,listings}
\usepackage{setspace}
\usepackage[font=small,labelfont=bf]{caption}
\usepackage{graphicx,subcaption}
\usepackage{algorithm,algpseudocode}
\usepackage{url}
\usepackage{cite}
\usepackage[hidelinks]{hyperref}
\usepackage{wrapfig}
\usepackage{bbding}
\usepackage{booktabs}       % professional-quality tables
\usepackage{nicefrac}       % compact symbols for 1/2, etc.
\usepackage{microtype}      % microtypography
\usepackage{amsmath,amssymb,amsfonts}
\usepackage{dsfont}
\usepackage{pifont}
\usepackage{enumitem}
\usepackage{etoolbox}
\usepackage[subtle]{savetrees}

\usepackage{inconsolata}

\newcommand{\minihead}[1]{{\vspace{.45em}\noindent\textbf{#1.} }}
\newcommand{\miniheadit}[1]{{\vspace{.45em}\noindent\textit{#1.} }}
\newcommand{\miniheadnopd}[1]{{\vspace{.45em}\noindent\textbf{#1} }}

\newcommand{\sntext}{NOSCOPE\xspace}
\newcommand{\sn}{\textsc{NoScope}\xspace}

\newcommand{\specialcell}[2][c]{%
    \begin{tabular}[#1]{@{}c@{}}#2\end{tabular}}

\newtoggle{rcolors} \togglefalse{rcolors}

\newcommand{\colora}[1]{\iftoggle{rcolors}{{\color{red}{#1}}}{#1}}

\newenvironment{tightcenter}{%
  \setlength\topsep{0pt}
  \setlength\parskip{0pt}
  \begin{center}
}{%
  \end{center}
}

\begin{document}

\newfont{\emailaddr}{phvr at 9pt}
\title{NoScope: Optimizing Neural Network Queries\\ over Video at Scale}

\author{Daniel Kang, John Emmons, Firas Abuzaid, Peter Bailis, Matei Zaharia\\ [1.45mm]    \affaddr{Stanford InfoLab}\\[1mm] {\emailaddr{noscope@cs.stanford.edu}}}

\maketitle

%!TEX root=vuse.tex
\begin{abstract}
  Recent advances in computer vision---in the form of deep neural
  networks---have made it possible to query increasing volumes of
  video data with high accuracy.  However, neural network inference is
  computationally expensive at scale: applying a state-of-the-art
  object detector in real time (i.e., 30+ frames per second) to a
  single video requires a \$4000 GPU. In response, we present \sn, a
  system for querying videos that can reduce the cost of neural
  network video analysis by up to three orders of magnitude via
  \emph{inference-optimized model search}. Given a target video, object to
  detect, and reference neural network, \sn automatically searches for
  and trains a sequence, or cascade, of models that preserves the
  accuracy of the reference network but is specialized to the target
  video and are therefore far less computationally expensive. \sn
  cascades two types of models: \emph{specialized models} that forego
  the full generality of the reference model but faithfully mimic its
  behavior for the target video and object; and \emph{difference
    detectors} that highlight temporal differences across frames. We
  show that the optimal cascade architecture differs across videos and
  objects, so \sn uses an efficient cost-based optimizer to search
  across models and cascades.  With this approach, \sn achieves two to
  three order of magnitude speed-ups (265-15,500$\times$ real-time) on
  binary classification tasks over fixed-angle webcam and surveillance
  video while maintaining accuracy within 1-5\% of state-of-the-art
  neural networks.
\end{abstract}

%!TEX root=vuse.tex
\section{Introduction}
\label{sec:intro}

Video represents a rich source of high-value, high-volume data: video comprised
over 70\% of all Internet traffic~\cite{cisco2016internet} in 2015 and over
300 hours of video are uploaded to YouTube every minute~\cite{youtube}. We can
leverage this video data to answer queries about the physical world, our lives
and relationships, and our evolving society.

\begin{figure}[t!]
  \centering
  \includegraphics[width=\columnwidth]{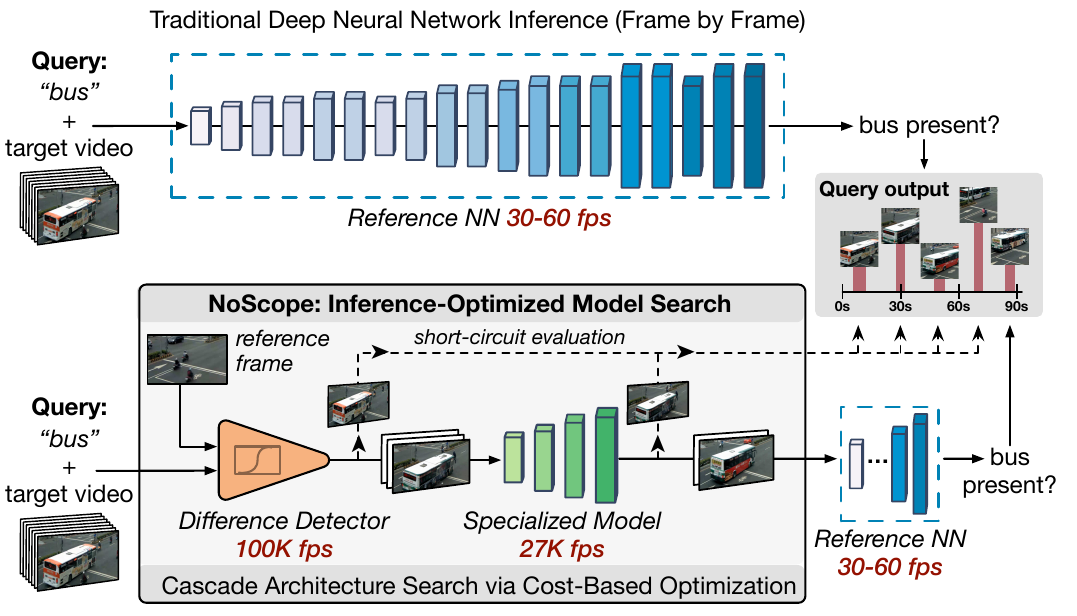}\vspace{1em}
  \caption{\sn is a system for accelerating neural network analysis
    over videos via inference-optimized model search. Given an input
    video, target object, and reference neural network, \sn
    automatically searches for and trains a cascade of
    models---including difference detectors and specialized
    networks---that can reproduce the binarized outputs of the
    reference network with high accuracy---but up to three orders of
    magnitude faster.}\vspace{-1.5em}
  \label{fig:overall}
\end{figure}

It is increasingly infeasible---both too costly and too slow---to rely
on manual, human-based inspection of large-scale video data. Thus,
automated analysis is critical to answering these queries at scale. The
literature offers a wealth of proposals for storing and
querying~\cite{video-db1,vdb-index,video-mining,video-db2,aref-vid,qbic,jacob-db,av-db,video-db3,mm-survey}
videos, largely based on classical computer vision techniques.
In recent times, however, deep
\emph{neural networks
(NNs)}~\cite{dl-nature,dl-book,malik60technical,ms-deep,wired-adoption}
have largely displaced classical computer vision methods due to their
incredible accuracy---often rivaling or exceeding human
capabilities---in visual analyses ranging from object
classification~\cite{russakovsky2015imagenet} to image-based cancer
diagnosis~\cite{thrun,re-cancer}.

%Automated analysis is critical to answering these queries at scale,
%and analysis capabilities are rapidly improving. It is increasingly
%infeasible---both too costly and too slow---to rely on manual,
%human-based inspection of large-scale video data. In addition, the
%literature offers a wealth of proposals for storing and
%querying~\cite{video-db1,vdb-index,video-mining,video-db2,aref-vid,qbic,jacob-db,av-db,video-db3,mm-survey}
%videos, largely based on classical computer vision
%techniques. However, recently, deep \emph{neural networks
%  (NNs)}~\cite{dl-nature,dl-book,malik60technical,ms-deep,wired-adoption}
%have largely displaced classical computer vision methods due to their
%incredible performance---often rivaling or exceeding human
%capabilities---in visual analyses ranging from object
%classification~\cite{russakovsky2015imagenet} to image-based cancer
%diagnosis~\cite{thrun,re-cancer}.

Unfortunately, applying NNs to video data is prohibitively expensive at
scale. The fastest NNs for accurate object detection run at 30-80
frames per second (fps), or 1-2.5$\times$ real time (e.g., 50 fps on an NVIDIA K80 GPU,
\textasciitilde\$4000 retail, \$0.70-0.90 per hour on cloud; 80 fps on an NVIDIA
P100, \textasciitilde\$4600 retail)~\cite{yolo9000,yolo,rcnn}.\footnote{In
  this work, we focus on multi-scale object detection, or identifying
  objects regardless of their scale in the
  image~\cite{yolo9000,yolo,rcnn}. Object detection models are
  more costly than classification models, which process images
  pre-centered on an object of interest, but are required to find
  objects in most realistic video applications.}
%That is, the many layers neurons that comprise a
%CNN require billions of floating point computations to evaluate.
% https://www.digikey.com/product-detail/en/omnivision-technologies-inc/OV07676-H20A/OV07676-H20A-ND/5322969
Given continued decreases in image sensor costs (e.g., < \$0.65 for a
640x480 VGA CMOS sensor), the computational overheads of NNs lead to a
three order-of-magnitude imbalance between the cost of data
acquisition and the cost of data processing. Moreover,
state-of-the-art NNs continue to get deeper and more costly to
evaluate; for example, Google's winning NN in the 2014 ImageNet
competition had 22 layers; Microsoft's winning NN in 2015 had 152
layers~\cite{ms-deep}.

In response, we present \sn, a system for querying videos that can
reduce the cost of NN-based analysis by up to three orders of
magnitude via \emph{inference-optimized model search}.
Our \sn prototype supports queries in the form of the presence or absence of a particular object class.
Given a query
consisting of a target video, object to detect, and reference \emph{pre-trained} neural
network (e.g., webcam in Taipei, buses, YOLOv2~\cite{yolo}), \sn automatically searches for and trains a
sequence, or \emph{cascade}~\cite{viola-jones}, of models that
preserves the accuracy of the reference network but are specialized to
the target query and are therefore far less computationally
expensive. That is, instead of simply running the reference NN over
the target video, \sn searches for, learns, and executes a
query-specific pipeline of cheaper models that approximates the
reference model to a specified target accuracy. \sn's query-specific
pipelines forego the generality of the reference NN---that is, \sn's
cascades are \emph{only} accurate in detecting the target object in
the target video---but in turn execute up to three orders of magnitude
faster (i.e., 265-15,500$\times$ real-time) with 1-5\% loss in
accuracy for binary detection tasks over real-world fixed-angle webcam and
surveillance video. To do so, \sn leverages both new types of models and
a new optimizer for model search:

First, \sn's \emph{specialized models} forego the full generality of
the reference NN but faithfully mimic its behavior for the target
query. In the context of our example query of detecting buses,
consider the following buses that appeared in a public webcam in
Taipei:\vspace{.05em}
\begin{tightcenter}
\hspace*{1.2em}
\includegraphics[width=.15\columnwidth]{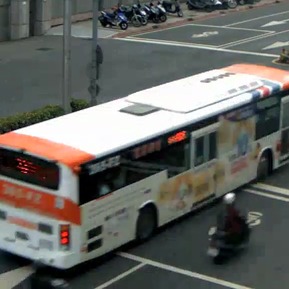}\hfill
\includegraphics[width=.15\columnwidth]{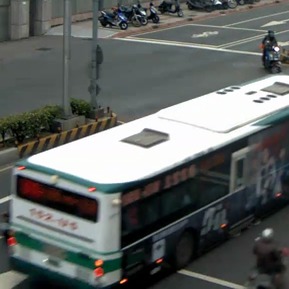}\hfill
\includegraphics[width=.15\columnwidth]{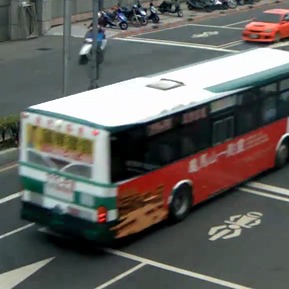}\hfill
\includegraphics[width=.15\columnwidth]{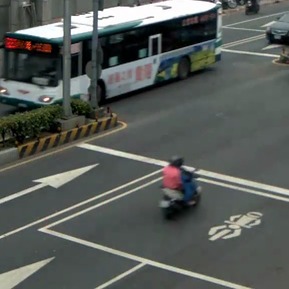}\hfill
\includegraphics[width=.15\columnwidth]{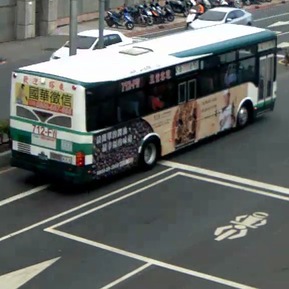}
\hspace*{1.2em}
\end{tightcenter}\vspace{-.02em}
In this video stream, buses only appear from a small set of
perspectives. In contrast, NNs are often trained to recognize
thousands of objects, from sheep to apples, and from different angles;
this leads to unnecessary computational overhead. Thus, \sn instead
performs \emph{model specialization}, using the full NN to generate
labeled training data (i.e., examples) and subsequently training
smaller NNs that are tailored to a given video stream and to a smaller
class of objects. \sn then executes these specialized models, which
are up to 340$\times$ faster than the full NN, and consults the full
NN only when the specialized models are uncertain (i.e., produce
results with confidence below an automatically learned threshold).

Second, \sn's \emph{difference detectors} highlight temporal
differences across frames. Consider the following frames, which
appeared sequentially in our Taipei webcam:\vspace{.05em}
\begin{tightcenter}
\hspace*{1.2em}
\includegraphics[width=.15\columnwidth]{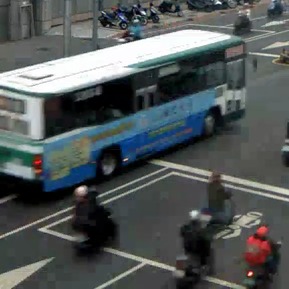}\hfill
\includegraphics[width=.15\columnwidth]{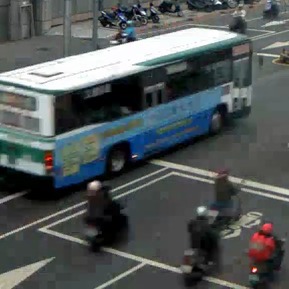}\hfill
\includegraphics[width=.15\columnwidth]{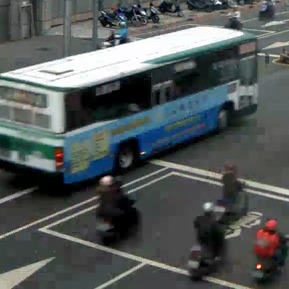}\hfill
\includegraphics[width=.15\columnwidth]{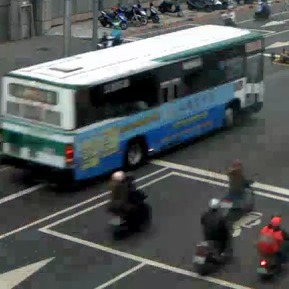}\hfill
\includegraphics[width=.15\columnwidth]{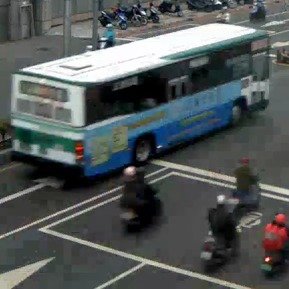}
\hspace*{1.2em}
\end{tightcenter}\vspace{-.02em}
These frames are nearly identical, and all contain the same bus.
Therefore, instead of running the full NN (or a specialized NN) on each
frame, \sn learns a low-cost difference detector (based on
differences of frame content) that determines whether the contents have
changed across frames.  \sn's difference detectors are fast and
accurate---up to 100k frames per second on the CPU.

A key challenge in combining the above insights and models is that the
optimal choice of cascade is data-dependent. Individual model
performance varies across videos, with distinct trade-offs between
speed, selectivity, and accuracy.  For example, a difference detector
based on subtraction from the previous frame might work well on mostly
static scenes but may add overhead in a video overseeing a busy
highway. Likewise, the complexity (e.g., number of layers) of
specialized NNs required to recognize different object
classes varies widely based on both the target object and video. Even
setting the thresholds in the cascade represents trade-off:
should we make a difference detector's threshold less aggressive to
reduce its false negative rate, or should we make it more aggressive
to eliminate more frames early in the pipeline and avoid calling a
more expensive model?

To solve this problem, \sn performs inference-optimized model search
using a cost-based optimizer that automatically finds a fast
model cascade for a given query and accuracy target.  The optimizer applies
candidate models to training data, then computes the optimal
thresholds for each combination of models using an efficient linear
parameter sweep through the space of feasible thresholds. The entire
search requires time comparable to labeling the sample data using the
reference NN (an unavoidable step in obtaining such data).

We evaluate a \sn prototype on binary classification tasks \colora{on
  cameras that are in a fixed location and at a fixed angle; this
  includes pedestrian and automotive detection as found in monitoring
  and surveillance applications.}  \sn demonstrates up to three order
of magnitude speedups over general-purpose state-of-the-art NNs while
retaining high---and configurable---accuracy (within 1-5\%) across a
range of videos, indicating a promising new strategy for efficient
inference and analysis of video data. In summary, we make the following contributions in this work:
\begin{enumerate}[itemsep=.1em,parsep=.4em,topsep=.5em]
\item \sn, a system for accelerating neural network queries over video
  via inference-optimized model search.

\item New techniques for $a)$ neural network model specialization
  based on a given video and query; $b)$ fast difference detection
  across frames; and $c)$ cost-based optimization to automatically
  identify the fastest cascade for a given accuracy target.

\item An evaluation of \sn on fixed-angle binary classification
  demonstrating up to three orders of magnitude speedups on real-world data.

\end{enumerate}

The remainder of this paper proceeds as
follows. Section~\ref{sec:background} provides additional background
on NNs and our target environment. Section~\ref{sec:architecture}
describes the \sn architecture. Section~\ref{sec:specialized}
describes \sn's use of model specialization, Section~\ref{sec:dds}
describes \sn's difference detectors, and Section~\ref{sec:optimizer}
describes \sn's inference-optimized model search via cost-based
optimization. Section~\ref{sec:implementation} describes the \sn
prototype implementation and Section~\ref{sec:limitations} describes
limitations of the current system. Section~\ref{sec:evaluation}
experimentally evaluates \sn, Section~\ref{sec:relatedwork} discusses
related work, and Section~\ref{sec:conclusion} concludes.

%!TEX root=vuse.tex
\section{Background}
\label{sec:background}

Given an input image or video, a host of computer vision methods can
extract valuable semantic information about objects and their
occurrences. In this section, we provide background on these methods,
focusing on object detection tasks: given an image, what objects does
it contain?  Readers familiar with computer vision may wish to proceed
to the next section.

\minihead{Object Detection History and Goals} Automated object
detection, or the task of extracting object occurrences and their
locations from image data, dates to at least the
1960s~\cite{summer-vision}. Classic techniques~\cite{cv-book,
  face-survey, classify-survey} combine machine learning methods such
as classification and clustering with image-specific featurization
techniques such as SIFT~\cite{sift}. \colora{More recent and advanced
  methods such as HOG~\cite{zhu2006fast}, deformable parts
  model~\cite{felzenszwalb2010object} and selective search
  \cite{uijlings2013selective} are among the most sophisticated of
  these classic approaches.}

Following these early successes, artificial neural networks have improved in
accuracy to near-human or better-than-human levels in the past five years.
Now, these ``deep'' models (with millions to billions of parameters) have
become not only feasible but also the preferred method for computer vision
tasks spanning image classification~\cite{russakovsky2015imagenet}, pedestrian
detection~\cite{ped-detect}, and automatic captioning~\cite{donahue}. To
understand why, consider the PASCAL VOC 2012~\cite{pascal-voc-2012}
leaderboard, in which classical methods were employed: the top three methods
(in accuracy) were NNs, and the winning entry, YOLOv2, runs at 80 fps. In
comparison, the top three classical methods take several seconds per image to
run and are 20\% less accurate. NNs power image processing tasks at online
services including Google, Facebook, Instagram, and Amazon as well as
real-world tasks including autonomous driving.

\minihead{NN Architecture} A \emph{neural network}~\cite{dl-book}
consists of a series of connected \emph{layers} that can process a
high-dimensional input image and output a simpler representation. Each
layer of a \emph{convolutional} NN corresponds to a step in
computation: these layers include \emph{convolutional} layers (that ``combine''
nearby pixels via convolution operators), \emph{pooling} layers (that
reduce the dimensionality of the subsequent layers), \emph{rectified
  linear unit (ReLU)} layers (that perform a non-linear
transformation), and a \emph{fully connected} layer (that outputs the
actual prediction based on prior layers). As illustrated
below~\cite{cnn-img}, combining multiple such layers in a ``deep''
architecture and by fitting the appropriate weights of the computation
(i.e., the ``neurons'') between stages, NNs can learn the structure
of common objects and detect their presence and absence in a given
image.
\begin{tightcenter}
\includegraphics[width=0.7\columnwidth]{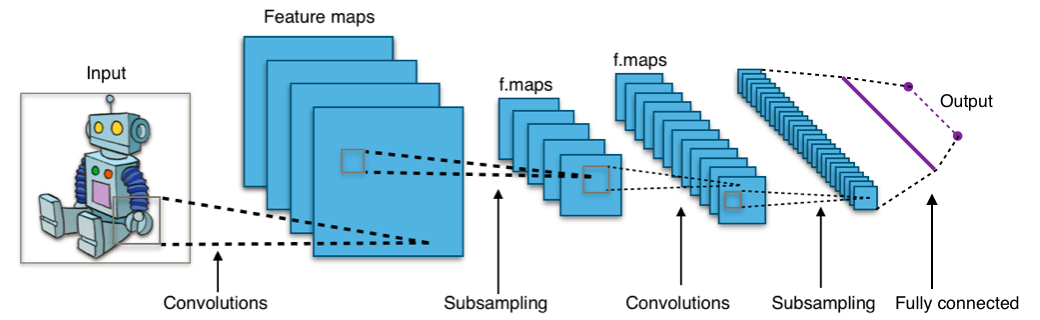}\\
\end{tightcenter}
Depending on the task, the best
model architecture (e.g., configuration of layers) varies, but the
overwhelming trend has been to build deeper models given more
training data.

\minihead{NN Training} Training NNs consists of fitting appropriate
weights to a given architecture such that the overall empirical error
on a given set of labeled training data is minimized. This process is
computationally expensive, but is now supported by a wide range
of software frameworks such as Google TensorFlow, Torch and Caffe.  To
train an object detector on video, we would first label a portion of a
video (or set of videos) by hand, marking which frames contained
people and which did not. Subsequently, we would feed each frame to a
NN training framework.  Conceptually, compared to prior computer
vision approaches that relied heavily on manually designed features
(e.g., gradients and edges~\cite{cv-book,sift}), NNs automatically
learn both low-level and high-level features from the training
dataset.  Due to the computational cost of training NNs, companies and
researchers have also published hundreds of pre-trained models, each
representing thousands of hours of CPU and GPU training time. From an
engineering perspective, these off-the-shelf, high-quality models are
easy to apply on new data, as we do (but faster) in \sn.

\minihead{NN Inference}
% cite: https://arxiv.org/pdf/1409.0575.pdf
Applying NNs to video---i.e., \emph{inference} on
video---consists almost exclusively of
passing individual video frames to a NN, one frame at a time. That is,
to detect objects in video, we evaluate the NN repeatedly, once per
frame. This is for several reasons. First, NNs are almost always
trained on static \emph{images}, not on moving video. Therefore,
evaluating them on video by converting them to a series of static
images is easy (if expensive). Second, historically, the resurgence of
interest in NNs has been driven by competitions such as ImageNet, in
which the only metric of interest is \emph{accuracy}, not inference
speed. The difference between the first- and second-place in ImageNet
2016's object recognition competition was 0.72\% (2.99\% vs. 3.71\%
top-5 classification error; in contrast, humans achieve approximately
5.1\% error).
Therefore, with few exceptions~\cite{kitti},
accelerated methods are not favored by the most prestigious
competitions in the field.  Finally, the handful of object recognition
CNNs that are optimized for inference time primarily use ``real time''
(i.e., 30 fps) as a target~\cite{yolo,yolo9000,rcnn},
aiming to evaluate one video in real time on a GPU.  In contrast, we
are interested in scaling to thousands of hours of video, possibly
from thousands of data feeds, for the purposes of large-scale video
classification; real time performance is not nearly fast enough for
our target scale.

%!TEX root=vuse.tex
\section{\sntext Architecture}
\label{sec:architecture}

\sn is a system for accelerating inference over video at scale. \sn
performs \emph{inference-optimized model search} to automatically identify
and learn a computationally efficient cascade, or pipeline of models,
that approximate the behavior of a reference neural network over a
given video and target object and to a desired accuracy. In this
section, we introduce the \sn system query interface and system
architecture.

\minihead{\sn Queries and Goal} In this work, we target binary
classification queries---i.e., presence or absence of a given class of
object in a video over time. In \sn, users input \emph{queries} by
selecting a target object class (e.g., one of the 9000 classes
recognized by YOLO9000, such as humans, cars, and buses~\cite{yolo})
as well as a target video. Subsequently, \sn outputs the time
intervals in the video when an object of the specified class was
visible according to a given \emph{reference model}, or full-scale NN
trained to recognize objects in images. \sn allows users to specify a
target accuracy in the form of false positive and false negative rates
and aims to maximize throughput subject to staying within these
rates.\footnote{A false positive is a case where \sn reports an
  object but running the reference model would have reported no object. A
  false negative is a case where \sn reports no object but the reference model
  would have reported one.} In summary, given these inputs, \sn's
goal is to \emph{produce the same classification output as applying
  the target model on all frames of the video, at a substantially
  lower computational cost and while staying within the specified
  accuracy target}.

%\begin{figure}[t]
%  \centering
%  \includegraphics[width=0.65\columnwidth]{figures/architecture_horizontal.pdf}
%  \caption{\sn design. \sn's optimizer selects a \emph{different}
%  configuration of specialized models and difference detectors for each video stream to
%  perform binary classification as quickly as possible without calling the full
%  target NN.}
%  \vspace{-1em}
%  \label{fig:arch1}
%\end{figure}

\minihead{System Components} \sn is comprised of three components, as
shown in Figure~\ref{fig:overall}: $a)$ specialized models, $b)$
difference detectors, and $c)$ an inference-optimized cost-based optimizer. When first
provided a new video, \sn applies the reference model to a subset of the
video, generating labeled examples. Using these examples, \sn searches
for and learns a cascade of cheaper models to accelerate the query on
the specific video. \sn subsequently runs the cascade over the
remainder of the video, stopping computation at the cheapest layer of
the cascade as soon as it is confident.

\sn uses two types of models.  First, \sn trains \emph{specialized
  models} (Section~\ref{sec:specialized}) that perform classification
tasks.  For example, while detecting humans with perfect accuracy in
all frames may require running the full reference model, we show that a much
smaller NN can output a confidence value $c$ that lets us safely label
a frame as ``no human'' if it is below some threshold
$c_\mathrm{low}$, label it as ``human'' if $c>c_\mathrm{high}$, and
pass the frame to the full NN if it is unsure (i.e., $c_\mathrm{low} < c < c_\mathrm{high}$).
Second, \sn uses \emph{difference detectors}
(Section~\ref{sec:dds}) to check whether the current frame is similar
to a recent frame whose label is known (e.g., for a camera looking at
a hallway, this could be an image where the hallway is empty).

Finally, to automatically search for and configure these models \sn
includes a \emph{cost-based optimizer} (Section~\ref{sec:optimizer})
that learns an efficient configuration of filters for each query to
achieve the target accuracy level (i.e., false positive and false
negative rates).  As discussed in Section~\ref{sec:intro}, we have
found (and empirically demonstrate) that customizing cascades for each
video is critical for performance.

In the next sections, we discuss each of these components in detail.
We begin with specialized models and difference detectors in
Sections~\ref{sec:specialized} and~\ref{sec:dds}. We then present our
cost-based optimizer in Section~\ref{sec:optimizer}.  Finally, we describe our
prototype implementation in Section~\ref{sec:implementation} and
current limitations in Section~\ref{sec:limitations}.

\section{Model Specialization} \label{sec:specialized}

The first key technique in \sn is the use of specialized models:
smaller models that faithfully mimic the behavior of a reference model
on a \emph{particular} task. Generic NNs can classify or detect
thousands of classes, and the generality of these methods naturally
leads to costly inference. Specialized models forego the full
generality of a generic, reference model but mimic its behavior on a
subset of tasks the generic model can perform. In query systems such
as \sn, we are generally only interested in identifying a small number
of objects---as opposed to the thousands of classes a generic NN can
classify---and, in video inference, such objects may only appear from
a small number of angles or configurations.

\sn performs model specialization by applying a larger, reference
model to a target video and using the output of the larger model to
train a smaller, specialized model. Given sufficient training data
from the reference model for a specific video, the specialized
model can be trained to mimic the reference model on the video while
requiring fewer computational resources (e.g., NN layers) compared to
the reference model. However, unlike the reference model, the
specialized model learns from examples from the target video and is
unlikely to \emph{generalize} to other videos or queries. Thus, by
sacrificing generalization and performing both training and inference on
a restricted task and input data distribution, we can substantially
reduce inference cost.

Critically, in contrast with related approaches to model
compression~\cite{jeff-distill,ba-distill,han-compression}, the goal
of model specialization is \emph{not} to provide a model that is
indistinguishable from the reference model on \emph{all tasks};
rather, the goal of model specialization is to provide a model that is
indistinguishable (to a given accuracy target) for a \emph{restricted}
set of tasks. This strategy allows efficient inference at the expense
of generality.

\sn uses shallow NNs as its specialized models. Shallow NNs have been
shown to be effective in other compression
routines~\cite{jeff-distill}, are efficient at inference time, and
naturally output a confidence in their classification.  \sn uses this
confidence to defer to the reference model when the specialized model
is not confident (e.g., when no loss in accuracy can be tolerated).

\sn implements specialized NNs based on the AlexNet
architecture~\cite{krizhevsky2012imagenet} (filter doubling, dropout),
using ReLU units for all the hidden layers and a softmax unit at the
end to return a confidence for the class we are querying. However, to
reduce inference time, \sn's networks are significantly shallower than
AlexNet.  As we discuss in Section~\ref{sec:optimizer}, \sn performs
automated model search by varying several parameters of the
specialized models, including the number of convolutional layers (2 or
4), number of convolution units in the base layer (32 or 64), and
number of neurons in the dense layer (32, 64, 128 or 256). As these
models provide different tradeoffs between speed and accuracy, \sn's
optimizer automatically searches for the best model for each video
stream and query.

Beyond configuring and learning a specialized model, \sn also selects
two thresholds on the specialized model's confidence $c$: $c_\mathrm{low}$ is the threshold at below which \sn outputs no object
in frame, and $c_\mathrm{high}$ is the threshold above which \sn outputs
object detected. For output values of $c$ between $c_\mathrm{low}$ and
$c_\mathrm{high}$, \sn calls the full reference NN on the frame.

The choice of threshold and the choice of model both determine speed
and accuracy. For example, a specialized NN with many layers and
convolution units might be much more accurate (resulting in a smaller
$[c_\mathrm{low}, c_\mathrm{high}]$) but more expensive to compute per
frame. In some cases, we should choose the less accurate NN that
passes more frames to our full model but is faster to evaluate on most
input frames; \sn's optimizer automates this decision.

To train specialized NNs, \sn uses standard NN training practices. \sn uses a
continuous section of video for training and cross-validation and learns NNs
using RMSprop~\cite{rmsprop} for 1-5 epochs, with early stopping if the
training loss increases. In addition, during model search, \sn uses a separate
evaluation set that is not part of the training and cross-validation sets for
each model.

As we illustrate in Section~\ref{sec:evaluation}, specialized models
trained using a large model such as YOLOv2 deliver substantial
speedups on many datasets. By appropriately setting $c_\mathrm{low}$
and $c_\mathrm{high}$, \sn can regularly eliminate 90\% of frames (and
sometimes all frames) without calling the full reference model while
still preserving its accuracy to a desired target.  We also show that
training these small models on \emph{scene-specific data} (frames from
the same video) leads to better performance than training them on
generic object detection datasets.

While we have evaluated model specialization in the context of binary
classification on video streams, ongoing work suggests this technique
is applicable to other tasks (e.g., bounding box
regression) and settings (e.g., generic image classification).

\section{Difference Detection}
\label{sec:dds}

\begin{figure}[t]
  \centering
  \begin{subfigure}[b]{0.32\columnwidth}
    \centering
    \includegraphics[width=0.95\columnwidth]{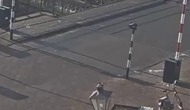}
    \caption{empty frame}
  \end{subfigure}
  \begin{subfigure}[b]{0.32\columnwidth}
    \centering
    \includegraphics[width=0.95\columnwidth]{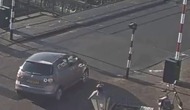}
    \caption{frame with a car}
  \end{subfigure} % \begin{subfigure}[b]{0.32\columnwidth} \centering
  \begin{subfigure}[b]{0.32\columnwidth}
    \centering
    \includegraphics[width=0.95\columnwidth]{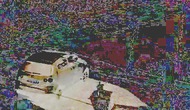}
    \caption{subtracted frames}
  \end{subfigure}
  \vspace{0.5em}
  \caption{Example of difference detection. The subtracted frame highlights
  the car that entered the scene.}
  %\vspace{-1em}
  \label{fig:diff_detect}
\end{figure}

The second key technique in \sn is the use of difference detectors:
extremely efficient models that detect changes in labels. Given a
labeled video frame (e.g. this frame does not have a car---a ``false''
in our binary classification setting) and an unlabeled frame, a
difference detector determines whether the unlabeled frame has the
same or different label compared to the labeled frame. Using
these difference detectors, \sn can quickly determine when video
contents have changed. In videos where the frame rate is much higher
than the label change rate (e.g., a 30 frame per second video
capturing people walking across a 36 foot crosswalk), difference
detectors can provide up to 90$\times$ speedups at inference
time.

In general, the problem of determining label changes is as
difficult as the binary classification task. However, as we have
hinted above, videos contain a high degree of temporal locality,
making the task of detecting redundant frames much
easier. Figure~\ref{fig:diff_detect} demonstrates this:
subtracting a frame containing an empty scene from a frame containing
a car distinctly highlights the car.
In addition, since \sn uses efficient difference
detectors (i.e., much more efficient than even specialized models),
only a small fraction of frames need to be filtered for difference
detectors to be worth the cost of evaluation.

\sn leverages two forms of difference detectors:
\begin{enumerate}[itemsep=.1em,parsep=.4em,topsep=.5em]
\item Difference detection against a fixed \emph{reference image} for
  the video stream, that contains no objects. For example, for a video of
  a sidewalk, the reference image might be a frame of an empty
  sidewalk. \sn computes the reference image by averaging frames where
  the reference model returns no labels.
  \item Difference detection against \emph{an earlier frame} a pre-configured
  time $t_\mathrm{diff}$ seconds into the past. In this case, if there are no
  significant differences, \sn returns the same labels that it output for
  the previous frame. \sn's optimizer learns $t_\mathrm{diff}$ based on the
  input data.
\end{enumerate}

The optimal choice of method is video-dependent, so \sn's optimizer performs
selection automatically.  For example, a video of a mostly empty
sidewalk might have a natural empty reference image that one can
cheaply and confidently compare with to skip many empty frames.  In
contrast, a video of a city park might always contain
mobile objects (e.g., people lying down in the grass), but the objects
might move slowly enough that comparing with frames 1 second ago can
still eliminate many calls to the expensive reference model.

Given the two frames to compare, \sn's difference detector computes
the Mean Square Error (MSE) between them as a measure of distance. \sn
either performs a comparison on the whole image, or a \emph{blocked}
comparison where it subdivides each image into a grid and computes the
metric on every grid block. In the blocked version, \sn then
trains a logistic regression (LR) classifier to weigh each block
when evaluating whether two images are different. The
blocked version is more expensive to compute, but it is useful when
part of the image does not contain relevant information. For example,
in a street view, the signal will change colors frequently but this is
not useful for detecting cars.

Using this distance metric, \sn determines whether the frame contents
have changed.  The appropriate \emph{firing threshold} for the metric,
or the difference in metric at which we say that the two frames
differ), which we will denote $\delta_\mathrm{diff}$, also depends on
the video and the query. For example, a lower MSE threshold might be
needed to detect light objects against a light background than dark
ones.

We also considered alternative metrics including Normalized Root Mean
Square Error, Peak Signal to Noise Ratio, and Sum of Absolute
Differences. While different methods performed differently,
MSE and blocked MSE were generally within a
few percent of the best method, so \sn currently only uses these two methods.

Finally, the difference detector has a configurable time
  interval $t_\mathrm{skip}$ that determines how often to perform a
  difference check.  For example, objects of interest may appear in a
  scene for much longer than a second, so \sn can test for differences
  every 15 frames in a 30 fps video.
  We refer to this behavior as frame skipping, and the frame skipping
  parameter directly creates a trade-off between accuracy and
  speed. Section~\ref{sec:filter_eval} demonstrates that frame
  skipping is effective but is not responsible for all performance
  gains offered by \sn.

As may be apparent, various configurations of the
difference detector will give varying tradeoffs between execution speed,
false positive rate and false negative rate depending on the video
and query in question. \sn's optimizer, described below, selects and adjusts the difference detector
automatically based on frames labeled using the expensive reference model to
navigate this trade-off.

\section{Cost-Based Model Search}
\label{sec:optimizer}

\sn combines model specialization and difference detectors using
\emph{inference-optimized model search}. \sn uses a cost-based
optimizer (CBO) to find a high-quality cascade of models (e.g., two vs four
layer specialized model) and thresholds ($\delta_\mathrm{diff}$,
$c_\mathrm{low}$, and $c_\mathrm{high}$).  The CBO takes as input a video as well as \emph{target accuracy values},
$\mathrm{FP}^*$ and $\mathrm{FN}^*$, for the false positive rate and
false negative rate, respectively.  Formally, the CBO solves the
following problem:
$$
\begin{array}{rl}
  \text{maximize} & E(\text{throughput}) \\
  \text{subject to} & \text{false positive rate} < \mathrm{FP}^* \\
         \text{and} & \text{false negative rate} < \mathrm{FN}^*
\end{array}
$$
This problem is similar to traditional cost-based estimation for
user-defined functions~\cite{udf-db} and streaming filter
processing~\cite{filters,filter-scheduling}, but \sn must also perform
cost estimation of each candidate model and set model-specific
parameters for each \emph{together} to achieve a target accuracy goal.
That is, \sn's problem is not simply a matter of ordering a set of
commutative filters, but also requires searching the model
architectures, learning the models, and setting their parameters
together to achieve an overall end-to-end cascade accuracy.

To solve this problem, \sn uses a combination of combinatorial search
for model architectures and efficient linear sweeps for the firing
thresholds $\delta_\mathrm{diff}$, $c_\mathrm{low}$, and
$c_\mathrm{high}$. \sn's CBO only considers plans involving up
to one difference detector and one specialized model because we found
that stacking more such models did not help significantly.  For such
plans, the CBO can search the \emph{complete} space of
combinations of difference detectors and specialized models at a
fraction of the cost of labeling all of the input video using a full
NN such as YOLOv2. We now describe the key components of the CBO.

\subsection{Training Data Generation}
\sn runs the full reference model on a subset of the video frames to
generate labeled data. For example, \sn could run the full NN over the
first few days of a new stream, and maintain this set using reservoir
sampling thereafter.  Given this labeled data, \sn's CBO begins
by splitting the input data into a \emph{training set} and an
\emph{evaluation set} that will only be used for model selection.

\subsection{Cost Model}
\label{sec:cost-model}

To determine which combination of difference detector and specialized
model to use, the CBO estimates the inference cost of each. \sn models execution time per frame
using the pass-through rates of the detector and specialized model in
question on the training data, and static measurements of the
execution time of each model on the target hardware.  For example,
consider a configuration with the following two filters:
\begin{enumerate}[itemsep=.1em,parsep=.4em,topsep=.5em]
  \item A Mean Squared Error (MSE) difference detector that is configured
  to check frames every $t_\mathrm{skip}$ seconds and fire if the frame's
  MSE with a reference image is higher than $\delta_\mathrm{diff}$.
  \item A two-layer NN with 64 convolutional units and 64 neurons in
  the dense layer, with detection thresholds
  $c_\mathrm{low}$ and $c_\mathrm{high}$.
\end{enumerate}

\sn's cost model determines the expected time per frame by first measuring
what fraction of frames remain after the skipping done by
$t_\mathrm{skip}$; call this fraction $f_s$.
Next, it computes the fraction of these frames in the training data
that pass the MSE filter, $f_m$.
Finally, it computes the fraction of these difference detector passing
frames whose confidence under the specialized NN lies between
$c_\mathrm{low}$ and $c_\mathrm{high}$ -- call this $f_c$.
The selectivities ($f_s$, $f_m$, and $f_c$) are estimated from a sample of the data.
The expected execution time per frame is then
$$
f_s T_\mathrm{MSE} + f_s f_m T_\mathrm{SpecializedNN} + f_s f_m f_c T_\mathrm{FullNN}
$$
where $T_\mathrm{MSE}$, $T_\mathrm{SpecializedNN}$ and $T_\mathrm{FullNN}$ are the
execution times per frame for the MSE filter, specialized NN and reference NN
respectively. These execution times are data-independent, so they can be measured
once per hardware platform.

\subsection{Model Search}

Given input data and the above cost model, \sn must find a suitable
cascade. There are three main challenges in finding an optimal
configuration. First, models may be complex and exhibit non-linearity
with respect to the cascade architecture and input parameters: their
selectivities will not be known a priori. Second, models are not
independent (cf.~\cite{filters}): one specialized model may complement
one difference detector but not another, and the viable combinations
of thresholds for each pair of models will differ across
scenes. Third, the full search space for model configurations is very
large because the firing thresholds $\delta_\mathrm{diff}$,
$c_\mathrm{low}$ and $c_\mathrm{high}$ are continuous values.

\sn addresses these challenges via a three-stage process.
First, it trains each filter individually on our training set.
Second, it examines each filter in isolation to determine
the scores for each frame.
Finally, it explores all combinations of difference detectors and
specialized models and use an efficient linear sweep of the viable
combinations of thresholds to determine the lowest-cost way to
combine these models.
In detail, \sn's search proceeds as follows:

\miniheadnopd{1.) Train filters:} First, \sn trains each
model (e.g., specialized NNs and LR based blocked difference
detectors) on the training data.  For the NNs, we consider a grid of
combinations of model architectures (e.g., 2 or 4 layers, 32 or 64
convolutional units, etc). Currently, this
is 24 distinct configurations.  When training these models,
\sn subdivides the original training data to create a
cross-validation set that is not part of training, for parameter selection.

\miniheadnopd{2.) Profile individual filters:} Next, \sn performs
selectivity estimation by profiling individual filters on its
evaluation set (the input data that the CBO set aside at the
beginning for evaluation) to determine their selectivity and
sensitivity.  These quantities are difficult to infer beforehand
because they depend significantly on the video, so \sn runs each
trained model on every frame of the evaluation set.  \sn then logs the
score that the filter gives to each frame (e.g., MSE or specialized NN
output confidence), which it uses to set the thresholds
$\delta_\mathrm{diff}$, $c_\mathrm{low}$ and $c_\mathrm{high}$.

\miniheadnopd{3.) Examine filter combinations:} Filter parameters are
\emph{not} independent because the thresholds for a pair of filters
may influence one another and therefore need to be set together.  For
example, if \sn uses MSE as our difference detection metric, the
threshold $\delta_\mathrm{diff}$ that it sets for difference detection
directly affects which \emph{frames} are passed to the downstream
specialized NN and which \emph{thresholds} $c_\mathrm{low}$ and
$c_\mathrm{high}$ it can set (e.g., if the difference detector induces
many false negatives, the specialized model must be more conservative).

\sn solves the problem of dependencies in filter selection via an
efficient algorithm to sweep feasible combinations of thresholds.
Specifically, for each difference detector $D$, \sn first sorts the
frames of the training data in decreasing order of the detector's
difference metric ($\delta$). This yields a list of frames $L_D$ where
we can easily sweep through firing thresholds
$\delta_\mathrm{diff}$. Next, for each prefix of the list $L_D$, \sn
computes the false positive and false negative rate for the given
$\delta_\mathrm{diff}$ threshold (i.e., frames that we will mislabel
due to the difference detector not firing).

Next, for each specialized NN, $C$, \sn sorts the unfiltered frames
by confidence. Given this list, we can set $c_\mathrm{low}$ and
$c_\mathrm{high}$ to obtain the desired false positive and false
negative rates $\mathrm{FP}^*$ and $\mathrm{FN}^*$: \sn begins with
thresholds at the extreme (0 and maximum confidence), then moves
$c_\mathrm{low}$ up until the false negative rate of the combined
difference detector and NN reaches $\mathrm{FN}^*$, and moves
$c_\mathrm{high}$ down until the combined false positive rate reaches
$\mathrm{FP}^*$. Finally, for each such combination of detector $D$,
NN $C$, $\delta_\mathrm{diff}$, $c_\mathrm{low}$, and
$c_\mathrm{high}$, \sn computes the expected throughput using our cost
model (Section~\ref{sec:cost-model}) and outputs the best
result. \colora{Note that because the CBO is only able
  to access training data, false positive and false negative rates are
  only guaranteed insofar as the training data reflects the testing
  data. Our experimental results demonstrate that this optimization
  strategy is sound for many real-world video feeds.}

\minihead{Running Time} The overall complexity of \sn's CBO
is $O(n_d n_c n_t)$, where $n_d$ is the total number of difference
detector configurations considered, $n_c$ is the total number of
specialized model configurations, and $n_t$ is the total number of
firing thresholds considered during the sweep down $L_D$. In addition,
we need to run each of the $n_d$ difference detectors and $n_c$
specialized model configurations on the training data once (but not
each \emph{pair of filters} together).  Overall, the running time of
the algorithm is often just a few seconds because each of $n_d$, $n_c$
and $n_t$ is small (less than 100).  Even training and testing the
specialized models we will test is faster than obtaining the ``ground
truth'' labels for the training data using a full-scale NN like
YOLOv2.  As we show in Section~\ref{sec:eval-train-time}, all steps of
the CBO together run faster than labeling hours-long videos
using a full NN.

%!TEX root=vuse.tex
\section{Implementation}
\label{sec:implementation}

We implemented a \sn prototype in C++ and TensorFlow (for the actual
inference tasks), and Python (for the \sn CBO). We optimized the
code by parallelizing many of the CPU-intensive operations over multiple
cores, using vectorized libraries, and batching data for
computations on the GPU; we provide our code as open
source\footnote{\texttt{https://github.com/stanford-futuredata/noscope}}.
These optimizations make a large difference in performance for many
configurations of \sn because the bulk of the video frames are
eliminated by the difference detector and specialized NN.

%We invested
%significant effort to achieve maximum performance for our difference
%detectors and specialized models, and provide our code as open
%source\footnote{https://github.com/stanford-futuredata/noscope}.  We
%parallelize many of the CPU-intensive operations over multiple cores,
%use vectorized libraries where possible, and batch data for
%computations on the GPU.  These optimizations make a large difference
%in performance for many configurations of \sn because the bulk of the
%video frames are eliminated by the difference detector and specialized
%NN.

At a high level, given a video, \sn performs the following four steps:
1) perform model search, 2) run the resulting difference detector, 3)
run the resulting specialized NN on the frames that pass the difference
detector, and 4) run the reference model such as YOLOv2 on the remaining
frames that are not confidently labeled by the specialized model.  We
next describe each of these steps in turn.

\minihead{Model Search Implementation} Our CBO is written in
Python, and calls our C++ code for difference detectors and NNs when
obtaining training data.  As described in Section~\ref{sec:optimizer},
the cost of the CBO itself is relatively small. The bulk of the
runtime is spent training the specialized NN models on its input data.

We used YOLOv2~\cite{yolo9000} as our reference model.
We pre-load the YOLOv2 NN parameters onto the GPU and keep it in memory across evaluations
to avoid reloading this large model both at optimization time and at inference time.

\sn loads video frames either directly from memory or from
OpenCV for video decoding. After loading frames, it resizes them for
downstream processing if needed using OpenCV (our NNs
require a specific size for input images, such as 416x416 pixels for YOLOv2).

\minihead{Difference Detectors}

We implemented difference detectors in OpenCV and C++. For the MSE
computation, we wrote hand-tuned C++ code to fuse together the required
operations: MSE requires computing
\texttt{sum$\left((a-b)^2\right)$}, where \texttt{a} and \texttt{b} are
two images, which would require materializing \texttt{a-b} in memory
using OpenCV's standard array operators. Finally, we parallelized the
difference detector at the level of frames---that is, we process
multiple decoded frames in parallel using different threads.

We used \texttt{scikit-learn} to train the logistic regression weights
for the blocked difference detectors that weigh different portions of
the image differently. We then evaluate the resulting logistic
regression models at runtime using C++ code.

\minihead{Specialized Models}
We train our specialized NNs using the TensorFlow framework.
As discussed in Section~\ref{sec:specialized}, we used standard practices
for NN training such as cross-validation and early stopping, and our CBO
selects which NN to run using an evaluation set that is distinct from the
training data.

During evaluation, we evaluate the specialized NNs on the GPU using TensorFlow's
C++ interface.
We batch input images before passing them to the GPU due to the high cost of communication.

Finally, a key pre-processing step is to mean-center the pixel values
and change the dynamic range in each color channel to [-1, 1].  We
implemented this step on the CPU using OpenCV and multithreading in a
similar manner to difference detection.

%!TEX root=vuse.tex
\section{Limitations}
\label{sec:limitations}

While \sn demonstrates significant promise in accelerating NN
inference, we wish to highlight several important limitations:

\minihead{Fixed-Angle Video} We designed the current prototype for the
task of object classification in fixed-angle video (e.g., traffic
cameras), which we believe represents an important subset of all video
data processing. Therefore, our current difference detectors only work
on video shot by static cameras, and our results for specialized CNNs
are also obtained on fixed-angle videos.  The video processing
community has also developed techniques to track objects when the
camera is moving~\cite{murray-active-camera}, so these techniques
present a promising approach to extend \sn to moving cameras.

\minihead{Binary Classification Task} We have only evaluated \sn on the
binary classification task; the techniques we present could already be
used for more complex boolean queries. We believe the same techniques
may apply elsewhere.

%We have only designed and evaluated \sn on binary classification of
%evaluated \sn on binary classification of objects in one object class
%from the target NN (e.g., finding all frames with people, or all
%frames with cars).  It would be natural to extend this type of
%optimization to more complex queries involving objects of multiple
%classes, localization (not just ``does the frame contain a person''
%but ``where is the person in the frame''), and counting queries (``how
%many people are in the frame'').  Some complex queries could be
%handled by ANDing or ORing the results of binary classification
%queries (e.g., ``does the frame contain a dog AND no human''), but
%others will require new forms of difference detectors and specialized
%models.

\minihead{Model Drift} Our current implementation assumes that the
training data obtained in \sn's optimizer is from the same
distribution as the subsequent video observed.  If the scene changes
dramatically after a certain point of time, \sn needs to be called
again to re-optimize for the new data distribution.  Likewise, if the
scene changes periodically during the video (e.g., by cycling between
day and night), the training data needs to contain frames
representative of all conditions.  We believe this is reasonable for
many fixed-angle camera scenarios, such as traffic or security
cameras, which observe a similar scene over a period of months or
years. For these types of videos, a sample of training data spread
throughout the year should be sufficient to produce good training data
for a model that can be applied continuously. However, it would be
interesting to address this limitation automatically (e.g., by
tracking model drift).
%We show in our Evaluation that just a few hours of training data can...

\minihead{Image Batching} Our implementation batches video frames for
greater efficiency. While acceptable for historical video that is
available all at once, batching can introduce a short delay for live
video, proportional to the number of frames batched.  For example,
batching together 100 frames at a time in a 30 fps video can add a
delay of 3.3 seconds. Alternatively, a system monitoring 100 streams
might be able to batch together data from different streams, applying
different NN models to each one, allowing a range of
optimizations~\cite{filter-sharing} we have not yet explored in the
context of streaming video.

%!TEX root=vuse.tex

\begin{table}[t!]
  \small
\centering
\setlength\itemsep{2em}
\caption{Video streams and object labels queried in our evaluation.}
\vspace{-1.1em}
\begin{tabular}{llcccc}
\specialcell{Video Name} & \specialcell{Object} & \specialcell{Resolution} & \specialcell{FPS} & \specialcell{\# Eval frames} & \specialcell{Length\\(hrs)} \\\hline
taipei        & bus      & 1000x570   & 30  & 1296k          & 12.0         \\
coral         & person   & 1280x720   & 30  & 1188k          & 11.0         \\
amsterdam     & car      & 680x420    & 30  & 1296k          & 12.0         \\
night-street  & car      & 1000x530   & 30  & 918k           & 8.5          \\
store         & person   & 1170x1080  & 30  & 559k           & 5.2          \\
elevator      & person   & 640x480    & 30  & 592k           & 5.5          \\
roundabout    & car      & 1280x720   & 25  & 731k           & 8.1          \\
\end{tabular}
\vspace{-0.1em}
\label{table:datasets}
\end{table}

\label{sec:end-to-end}
\begin{figure}[t!]
  \centering
  \begin{subfigure}[b]{0.45\columnwidth}
    \centering
    \includegraphics[width=0.8\columnwidth]{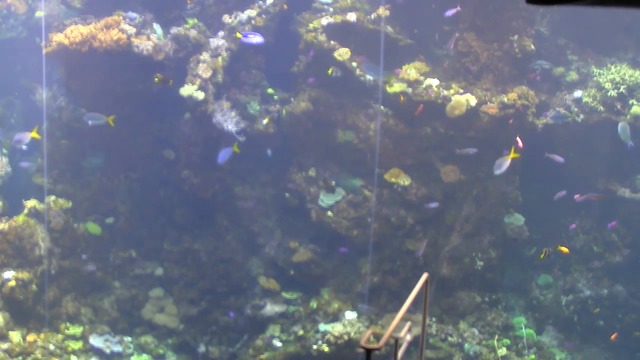}
    \caption{coral without a person}
  \end{subfigure}
  \begin{subfigure}[b]{0.45\columnwidth}
    \centering
    \includegraphics[width=0.8\columnwidth]{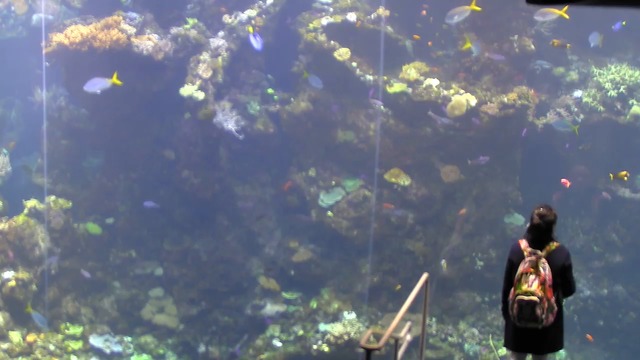}
    \caption{coral with a person}
  \end{subfigure}
  \vspace{0.5em}
  \caption{Example query, with and without the object of interest.}
  \vspace{-1em}
  \label{fig:screencaps}
\end{figure}

\section{Evaluation}
\label{sec:evaluation}

We evaluate \sn on binary classification for real-world webcam and
surveillance videos. We illustrate that:

\begin{enumerate}[itemsep=.1em,parsep=.4em,topsep=.5em]
\item \sn achieves 40-5400$\times$ higher throughput (i.e. frames processed per
  second) than state-of-the-art NN models while maintaining 99\% of
  their accuracy, and 100-10000$\times$ higher throughput while maintaining 90+\%
  accuracy (\S~\ref{sec:eval-e2e}).
\item The configuration of filters dramatically affects \sn's overall
  performance. Moreover, the optimal cascade for one video will almost
  always provide poor performance if used on another video
  (\S~\ref{sec:eval-optimizer}).
\item \sn's CBO consistently sets filter parameters
  that provide large performance improvements while maintaining
  specified accuracy. Difference detection and specialized models
  improve the throughput up to 3$\times$ and 340$\times$ respectively (\S~\ref{sec:eval-optimizer}).
\item Specializing models for a given video provides a 1.25-25$\times$
  improvement over training models for the same detection task across
  multiple environments (\S~\ref{sec:eval-model-spec}).
\item \sn's outperforms both classical computer vision baselines as well as
  small, binary-classification NNs that are not specialized to a given video (\S~\ref{sec:eval-baselines}).
\end{enumerate}

\subsection{Experimental Setup}
\label{sec:eval-setup}

\minihead{Evaluation Queries} We evaluate \sn on a fixed set of
queries depicted in Table~\ref{table:datasets}.
We use YOLOv2~\cite{yolo9000}, a state-of-the-art multi-scale NN,
as our reference model.
YOLOv2
operates on 416x416 pixel images (resizing larger or smaller images).
YOLOv2 achieves 80 fps on the Tesla P100 GPU
installed on our measurement machine. We obtain videos from seven
webcams---six from YouTube Live, and one that we manually obtained. We
split each video into two parts: training and evaluation. Five videos
have two days worth of video, with 8-12 hours of footage per day due
to lighting conditions; for these videos, we use the first day of
video for training and the second day for evaluation. For two videos,
we use the first 2.3 hours for training and separate an evaluation set
(5-8 hours) by a minimum of 30 minutes. Figure~\ref{fig:screencaps}
illustrates an example. By default, we set a target of 1\% false
positive and 1\% false negative rates.

\minihead{Evaluation Metrics} We measure throughput by timing the
complete end-to-end system excluding the time taken to decode video
frames. We omit video decode times for three reasons. First, it is
standard to omit video decode time in computer
vision~\cite{yolo9000,yolo,rcnn}. Second, both GPUs and CPUs can
decode video at a high rate because of built-in hardware
decoders~\cite{gpu-decoding-presentation}. For example, our Tesla P100
can decode 400x400 H.264 video at 4000 fps and 400x400 MPEG video at
5600 fps with a single hardware decoder; some processors have multiple
decoders. Third, for some visual sensors, we can directly obtain a raw
stream of frames from the hardware.

We measure accuracy by comparing frames labeled by the reference model and
\sn in 30 frame windows. For a window to be considered labeled
correctly, both systems must agree on the presence of the target
object in 28 of the 30 frames in a window. We use this somewhat
permissive accuracy measure because YOLOv2 often intermittently fails
to label objects that are clearly visible in the frame.

\minihead{Hardware Environment} We perform our experiments on an
NVIDIA DGX-1 server, using at most one Tesla P100 GPU and 32 Intel
Xeon E5-2698 v4 cores during each experiment. The complete system had
80 cores and multiple GPUs, but we limited our testing to a subset of
these so our results would be representative of a less costly
server. The system also had a total of 528 GB of RAM.

\subsection{End-to-End Performance}
\label{sec:eval-e2e}

\begin{figure}[t!]
  \captionsetup[subfigure]{aboveskip=-2pt}
  \centering
  \begin{subfigure}[b]{0.45\columnwidth}
    \centering
    \includegraphics[width=0.9\columnwidth]{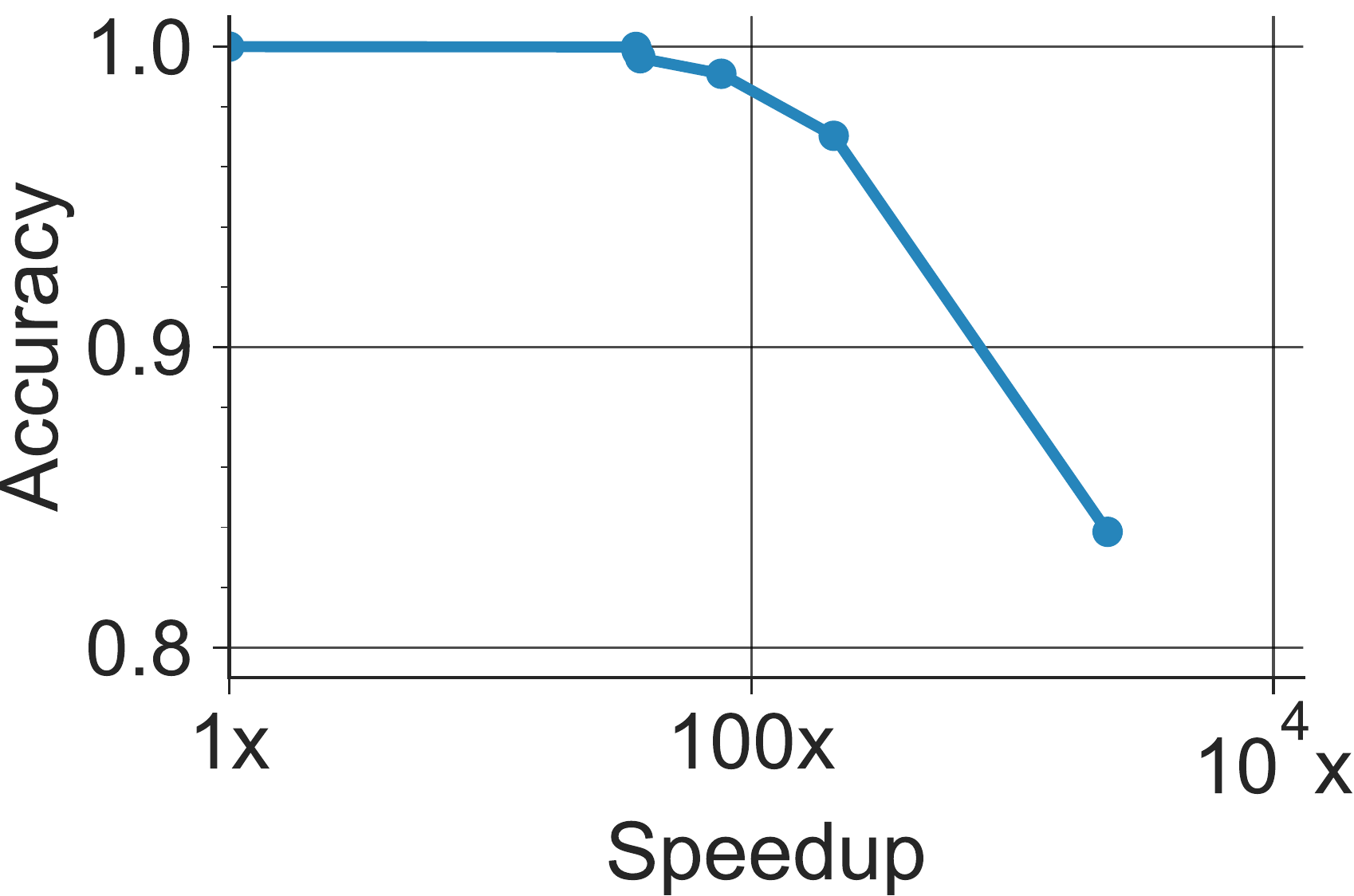}
    \vspace{0.4em}
    \caption{taipei}
  \end{subfigure}
  \begin{subfigure}[b]{0.45\columnwidth}
    \centering
    \includegraphics[width=0.9\columnwidth]{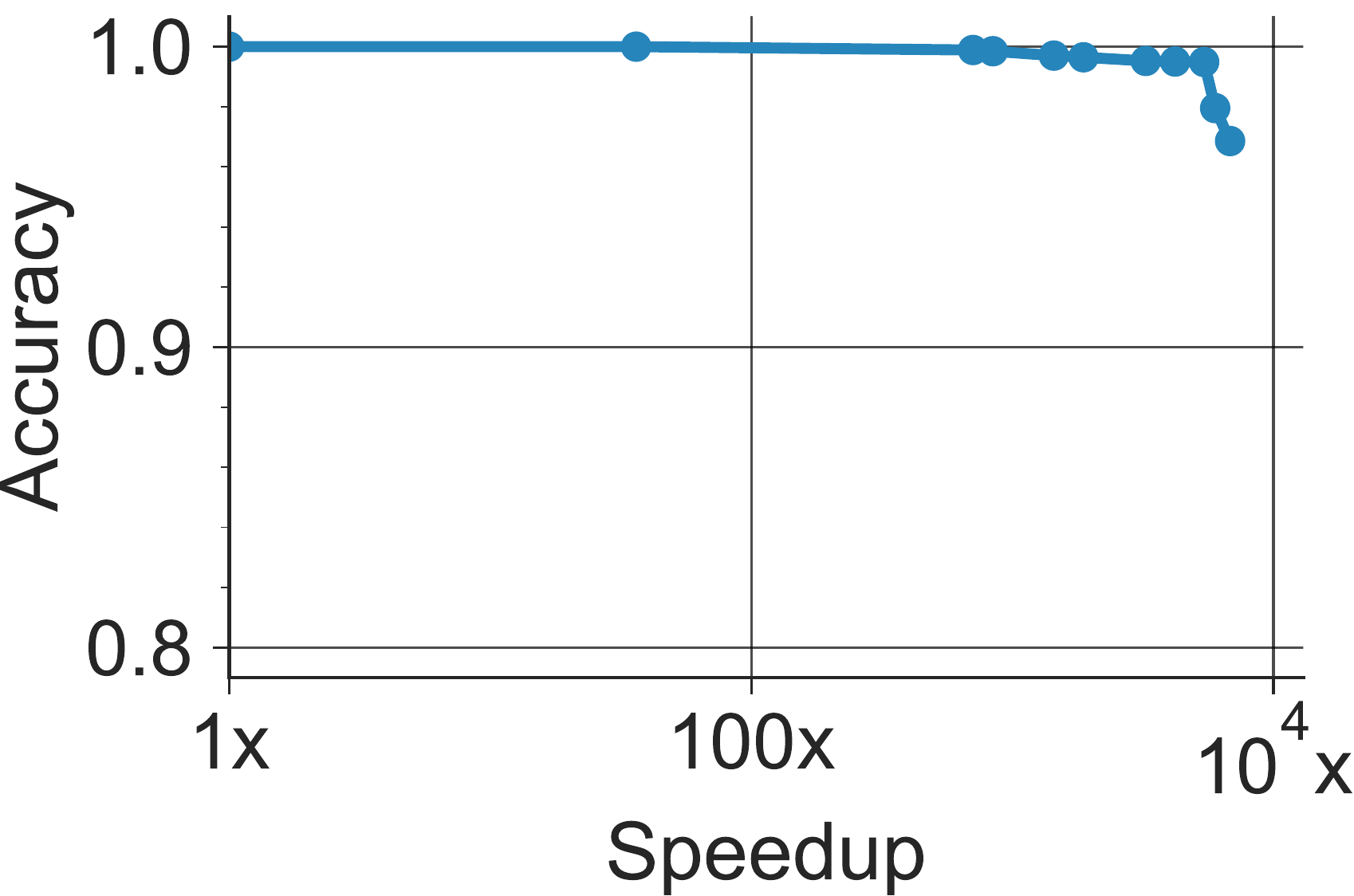}
    \vspace{0.4em}
    \caption{coral}
  \end{subfigure}
  \begin{subfigure}[b]{0.45\columnwidth}
    \centering
    \includegraphics[width=0.9\columnwidth]{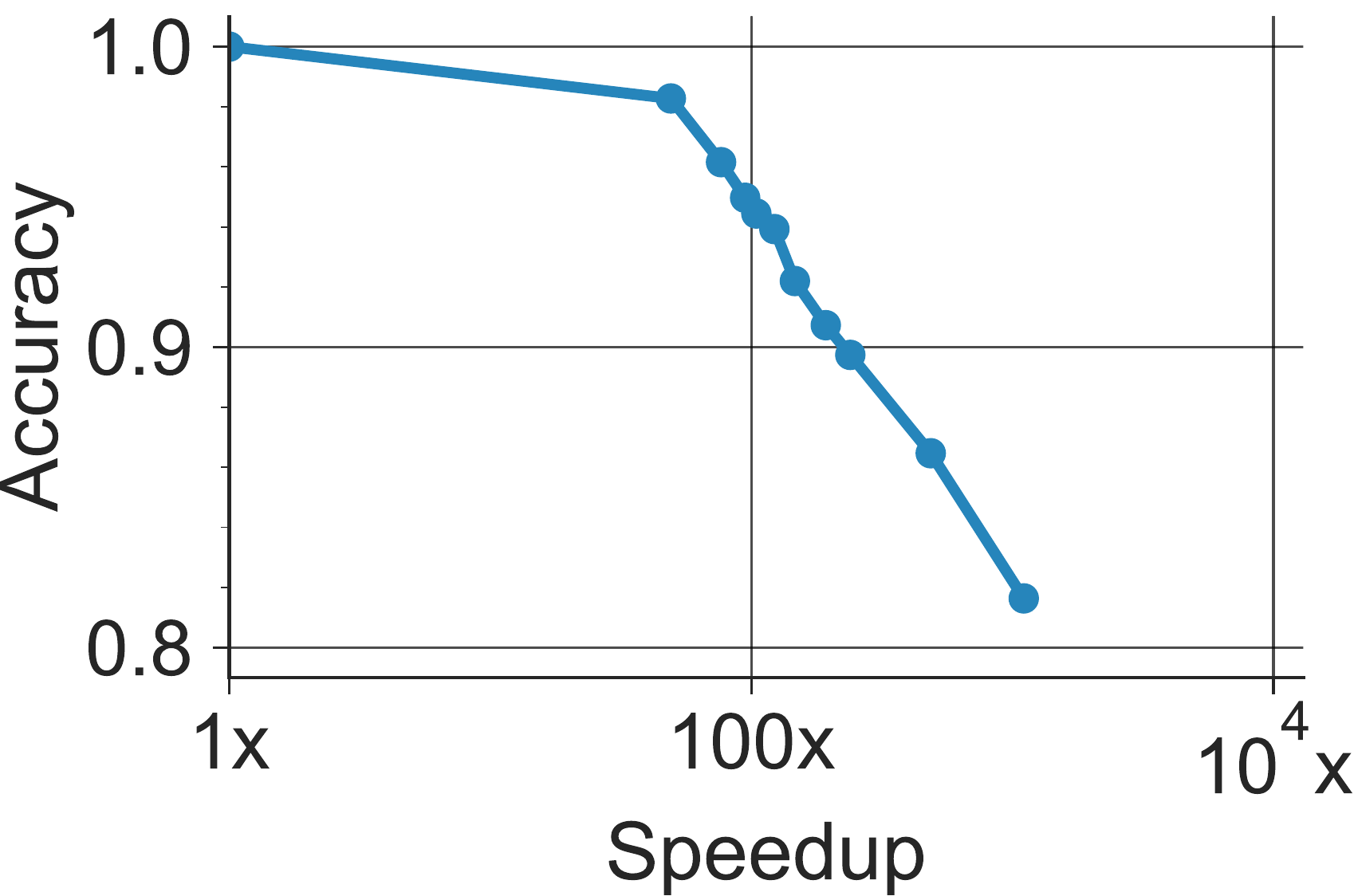}
    \vspace{0.4em}
    \caption{amsterdam}
  \end{subfigure}
  \begin{subfigure}[b]{0.45\columnwidth}
    \centering
    \includegraphics[width=0.9\columnwidth]{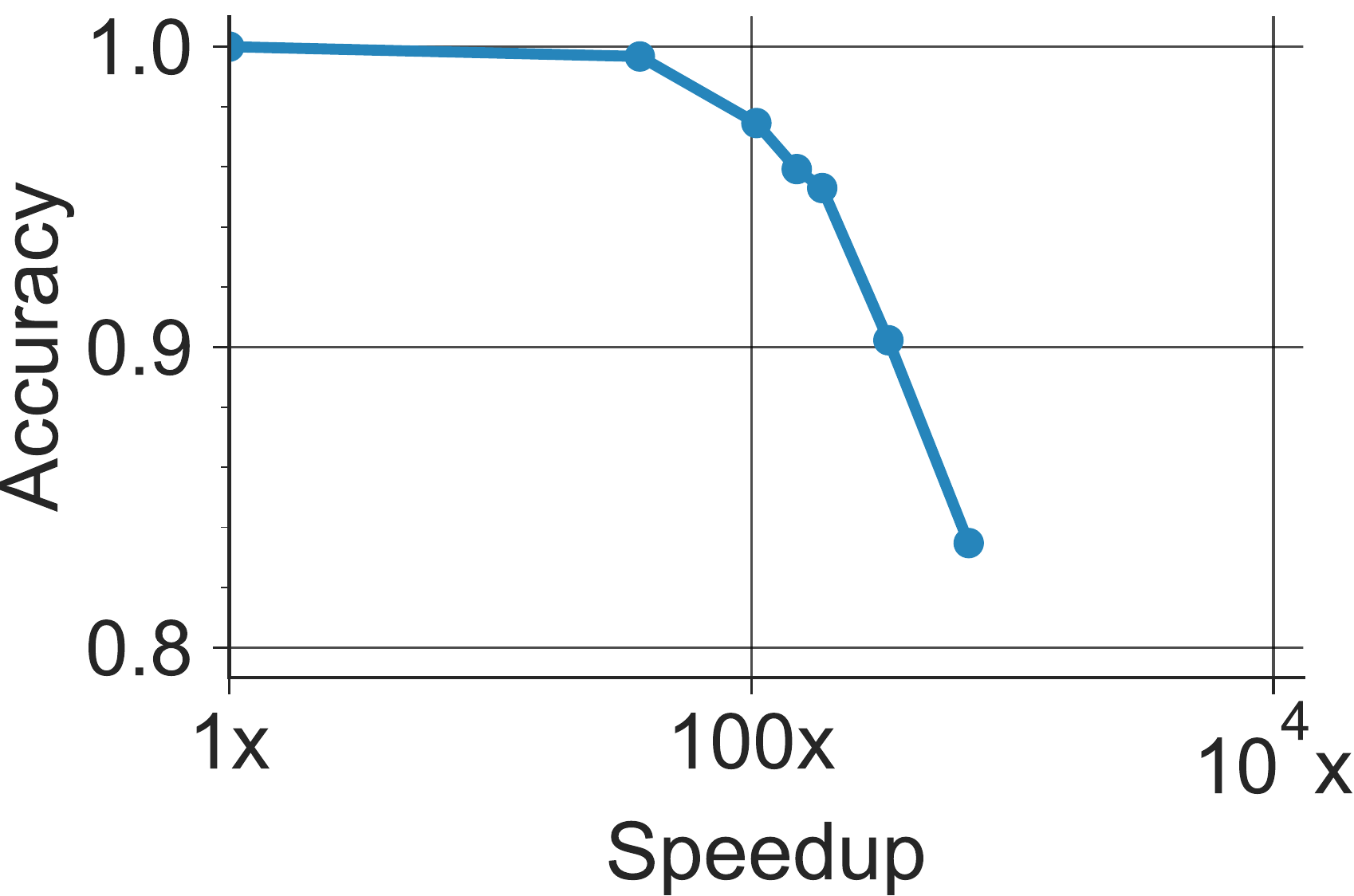}
    \vspace{0.4em}
    \caption{night-street}
  \end{subfigure}
  \begin{subfigure}[b]{0.45\columnwidth}
    \centering
    \includegraphics[width=0.9\columnwidth]{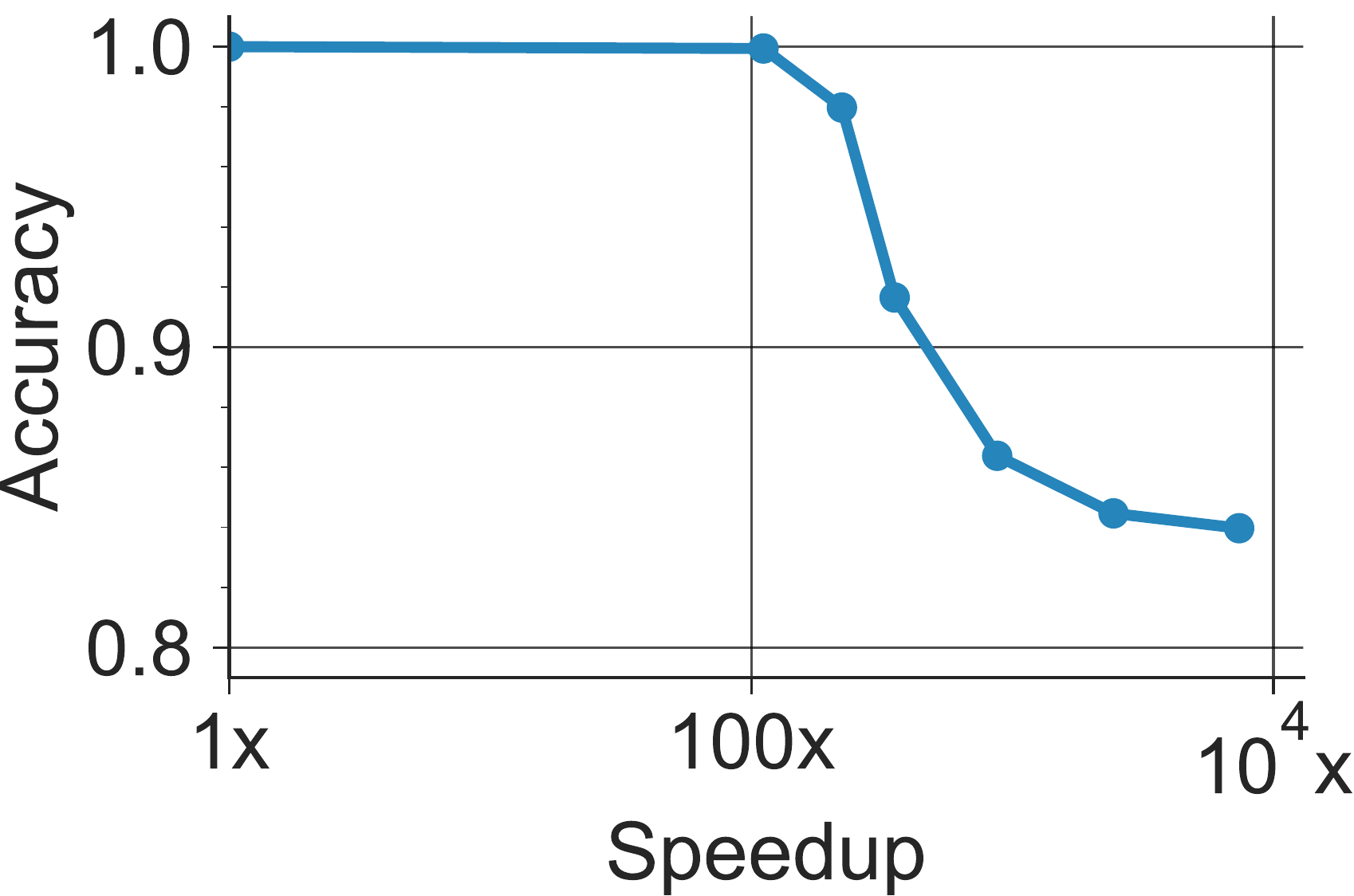}
    \vspace{0.4em}
    \caption{store}
  \end{subfigure}
  \begin{subfigure}[b]{0.45\columnwidth}
    \centering
    \includegraphics[width=0.9\columnwidth]{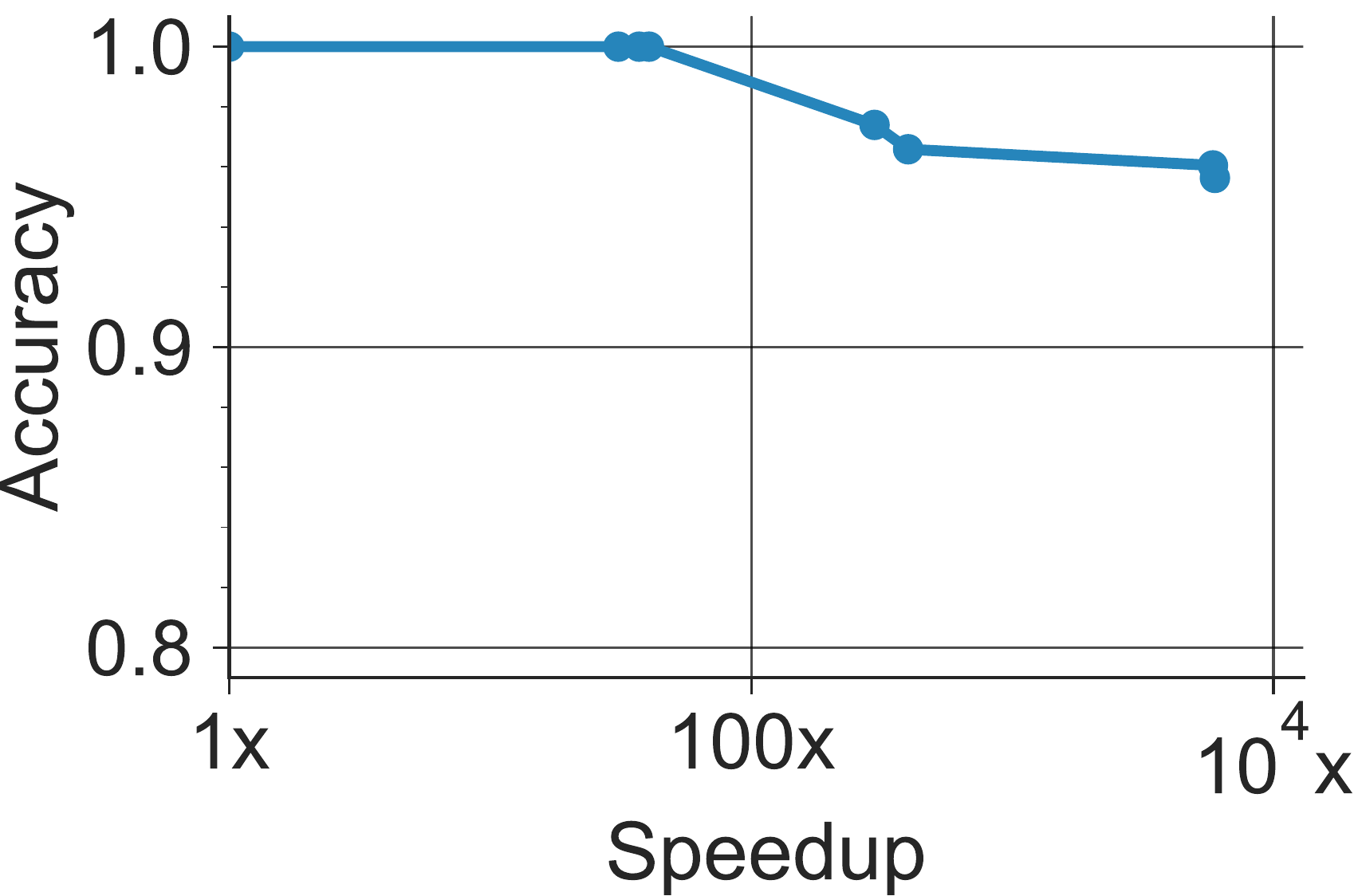}
    \vspace{0.4em}
    \caption{elevator}
  \end{subfigure}
  \begin{subfigure}[b]{0.45\columnwidth}
    \centering
    \includegraphics[width=0.9\columnwidth]{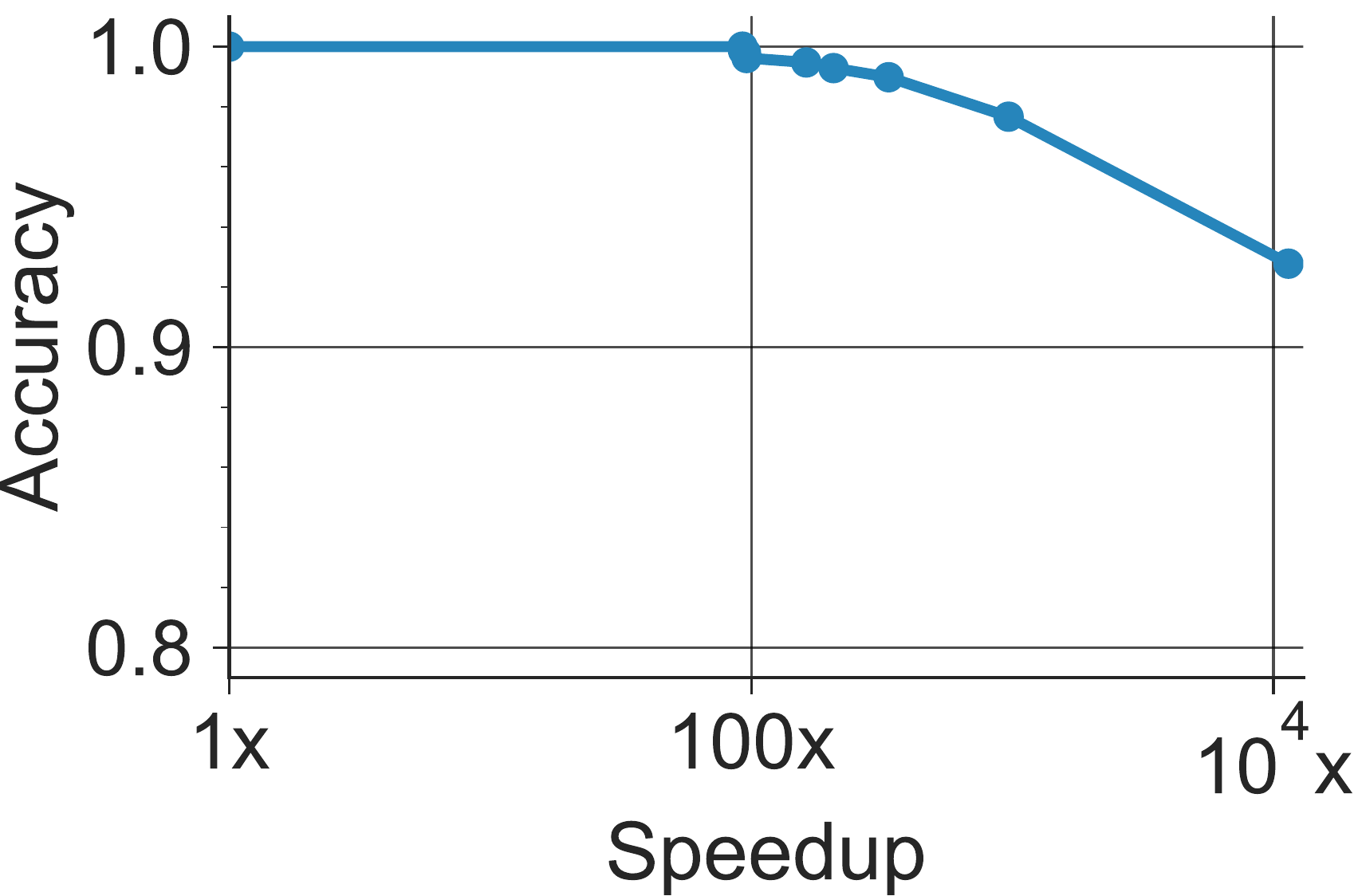}
    \vspace{0.4em}
    \caption{roundabout}
  \end{subfigure}
  \vspace{0.5em}
  \caption{Accuracy vs.~speedup achieved by \sn on each dataset.
  Accuracy is the percent of correctly labeled time windows, and speedup is over YOLOv2. Note the y-axis starts at 80\%.}
  \vspace{-0.5em}
  \label{fig:fu}
\end{figure}

Figure~\ref{fig:fu} illustrates the overall range of performance that
\sn achieves on our target queries. For each dataset, we obtained the
points in the plot by running \sn's CBO with increasing false
positive and false negative thresholds ($\mathrm{FP}^*$ and
$\mathrm{FN}^*$, with $\mathrm{FP}^* = \mathrm{FN}^*$) and measuring
the resulting speedup. \sn demonstrates several orders of magnitude
speedup across all the datasets, with magnitude depending on the
desired accuracy levels.  In all cases, \sn achieves a 30$\times$
speedup with at least 98\% accuracy. In many cases, this level of
accuracy is retained even at a 100$\times$ speedup, and \sn can obtain
1000$\times$ to 10,000$\times$ speedups at 90+\% accuracy.  The video
with the lowest peak speedup at the 90\% accuracy mark is
\texttt{taipei}, which shows a busy intersection---thus, the
difference detectors cannot eliminate many frames. However, even in
this video, \sn can offer an 30$\times$ speedup over YOLOv2 with no
loss in accuracy.

\begin{table}[t!]
\small
\centering
\setlength\itemsep{2em}
\caption{Filter types and thresholds chosen by \sn's CBO for each video at
1\% target false positive and false negative rates.
Both the filter types (e.g., global or blocked MSE) and their thresholds (e.g., the difference
in MSE that is considered significant, or the upper and lower detection thresholds for the
specialized models) vary significantly across videos. For the specialized models (denoted SM), L denotes
the number of layers, C the number of convolutional units, and D the dimension of the
dense layers.}
\vspace{-0.5em}
\begin{tabular}{lllllllll}
  \specialcell{Video\\Name} & \specialcell{\\DD} & \specialcell{\\\large$\delta_\mathrm{diff}$} & \specialcell{SM\\(L)} & \specialcell{SM\\(C)} & \specialcell{SM\\(D)} & \specialcell{\\ \large$c_\mathrm{low}$} & \specialcell{\\ \large$c_\mathrm{high}$} \\\hline
  taipei       & global   & 37.5   & 2 & 64 & 32   & 0.114  & 0.994  \\
  coral        & blocked  & 147.3  & 2 & 16 & 128  & 0.0061 & 0.998  \\
  amsterdam    & global   & 0.0019 & 2 & 64 & 256  & 0.128  & 0.998  \\
  night-street & global   & 0.441  & 2 & 16 & 128  & 2.2e-7 & 0.176  \\
  store        & blocked  & 336.7  & 2 & 32 & 128  & 0.010  & 0.998  \\
  elevator     & global   & 0.0383 & 2 & 32 & 256  & 0.004  & 0.517  \\
  roundabout   & global   & 0.115  & 4 & 32 & 32   & 0.0741 & 0.855  \\
  % Add new number below if time allows
  %% night-street    & global MSE & reference & 0.30  & MNIST & 2 &  64 & 128  & 1.27e-9 & 1.0  \\
  %%   store           & blocked MSE & previous & 336.7 & MNIST & 2 & 32 & 128  & 0.006 & 0.998  \\
  %%   roundabout      & global MSE & reference & 0.114 & CIFAR10 & 4 & 32 & 2  & 0.050 & 0.904  \\
  %%   coral           & blocked MSE & previous & 132.5 & CIFAR10 & 4 & 1 & 256  & 0.0249 & 1.0  \\
  %%   elevator        & blocked MSE & previous & 68.51 & MNIST & 2 & 32 & 256  & 0.002 & 1.0  \\
  %%   amsterdam       & global MSE & reference & 0.002 & MNIST & 2 & 64 & 256 & 0.113 & 0.998  \\
  %%   taipei          & global MSE & reference & 37.32 & MNIST & 2 & 64 & 32 & 0.069  & 0.997  \\
\end{tabular}
\vspace{0.5em}
\label{table:optimizer-settings}
\end{table}

\subsection{Impact of the CBO}
\label{sec:eval-optimizer}

To better understand the source of speedups, we explored the impact of \sn's CBO on performance.
We begin by showing that the filter types and thresholds chosen by the CBO differ significantly
across videos based on the characteristics of their contents.
We also show that choosing other settings for these parameters would greatly decrease speed or
accuracy in many cases, so parameters cannot be transferred across videos.

\minihead{Configurations Chosen Across Datasets}
Table~\ref{table:optimizer-settings} shows the difference detectors, specialized models,
and detection thresholds chosen by the CBO across our sample datasets for
a 1\% target false positive and false negative rate.
We observe that these are substantially different across videos, even when the CBO
selects the same filter class (e.g., difference detection based on global MSE).
We make a few observations about these results:

First, the best type of MSE chosen depend on the
video. For example, \texttt{coral} and \texttt{store} are
scenes with a dynamic background (e.g., \texttt{coral} shows an
aquarium with colorful fish swimming in the background, and \sn is
asked to detect people in the scene). In these scenes,
computing MSE against several frames past instead of against a single
``empty'' reference frame is more effective.

Second, the chosen thresholds also differ significantly. For example,
  \texttt{taipei} has a high difference detection threshold due to high levels of small-scale
  activity in the background, but the target objects, buses, are generally
  large and change more of the frame. The upper and lower thresholds
  for NNs also vary even across the same target object class, partly due to varying
  difficulty in detecting the objects in different scenes.
  For example, in some videos, the $c_\mathrm{low}$ threshold for declaring that there
  is no object in a frame is extremely low because increasing it would lead to too many
  false negatives.

Third, the best specialized model architectures also varied by video and query.
  In particular, we found that in many videos, the larger NNs (with 4 layers or
  with more convolutional units and dense layer neurons) would overfit given the
  fairly small training set we used (150,000 frames out of the 250,000 frames set
  aside for both training and evaluation). However, the best combination of the
  model architecture parameters varied across videos, and \sn's training routine that selects models
  based on an unseen evaluation set automatically chose an accurate model that did
  not overfit.

\minihead{Non-Transferability Across Datasets}
As the best filter configurations varied significantly across
datasets, transferring parameters from one dataset to another leads to
poor performance:

\begin{figure}[t!]
  \centering
  \includegraphics[width=0.99\columnwidth]{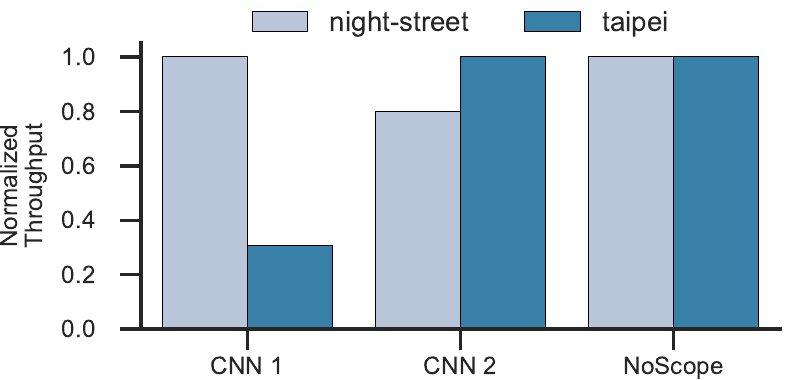}
  \vspace{0.5em}
  \caption{
  Normalized performance of \sn with two different specialized NN models on
  the \texttt{night-street} and \texttt{taipei} videos.
  We see that choosing a different NN architecture for each video, even though
  this architecture performed well on another dataset, reduces throughput.
  \sn automatically selects the best-performing NN.
  }
  \vspace{-1em}
  \label{fig:transfer-architectures}
\end{figure}

\miniheadit{Specialized Model Architectures} We used the specialized
model architecture from \texttt{night-street} (a NN with parameters
L=2, C=16, D=128) on the \texttt{taipei} dataset (whose optimal NN had
L=2, C=64, D=32), and vice-versa. We transferred \emph{only} the
\emph{architecture}, not the learned model from each dataset and
trained a new model with each architecture for the new dataset to
evaluate its performance there. Although these architectures have
similar properties (e.g., two layers), they required significantly
different parameters to achieve our 1\% target false positive and
false negative rates on each dataset.  This resulted in a 1.25$\times$
to 3$\times$ reduction in throughput for the overall \sn pipeline, as
depicted in Figure~\ref{fig:transfer-architectures}.
% TODO
%this reduction was caused primarily from the cutoffs for the specialized
%models being less aggressive and passing more frames to the full model.

\begin{figure}[t!]
  \centering
  \includegraphics[width=0.99\columnwidth]{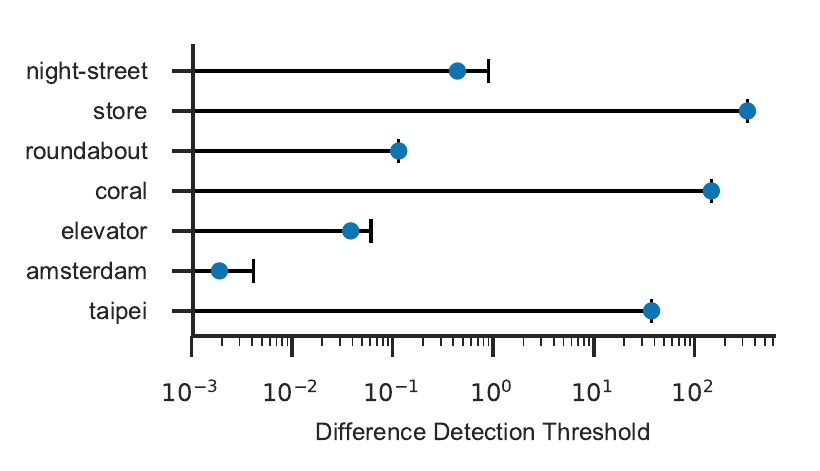}
  \caption{Firing thresholds $\delta_\mathrm{diff}$ chosen by the CBO for each
  video (blue dots), along with the range of thresholds that can achieve 1\% false
  positive and false negative rate for that video (black lines).
  }
  \vspace{-0.5em}
  \label{fig:dd-thresholds}
\end{figure}

\miniheadit{Detection Thresholds} We plotted the range of feasible
thresholds for the difference detector ($\delta_\mathrm{diff}$), as
well as the actual threshold chosen, in
Figure~\ref{fig:dd-thresholds}.  Feasible thresholds here are the ones
where the system can stay within the 1\% false positive and false
negative rates we had requested for the CBO.  Beyond a certain
upper limit, the difference detector alone will introduce too many
incorrect labels \colora{(on the validation set)}.  As we see in the
plot, the range of values for each video is different and the
best-performing threshold is often near the top of this range,
\colora{but not always.}  \colora{For example, \texttt{coral} is near
  the top, but \texttt{amsterdam} is lower, this is due to the
  downstream performance of the specialized models.}  Thus, attempting
to use a common threshold between videos would either result in
failing to achieve the accuracy target (if the threshold is too high)
or lower performance than the threshold \sn chose (if the threshold is
too low).

%\subsubsection{Effectiveness in Meeting Accuracy Targets}
%
%\begin{figure}[t!]
%  \centering
%  %
%  \begin{subfigure}[b]{0.49\columnwidth}
%    \centering
%    \includegraphics[width=\columnwidth]{figures/fu_plots/fp_plot.pdf}
%    \caption{False positive rate\\(i.e., spurious object detections).}
%  \end{subfigure}
%  %
%  \begin{subfigure}[b]{0.49\columnwidth}
%    \centering
%    \includegraphics[width=\columnwidth]{figures/fu_plots/fn_plot.pdf}
%    \caption{False negative rate\\(i.e., missed object detections).}
%  \end{subfigure}
%
%  \vspace{0.25cm}
%  \caption{The optimizer in \sn is given a target false positive and false
%    negative rate by the user. Using these constraint, the optimizer sets the
%    parameters of the various filter components to maximize the throughput on
%    the model evaluation set. These parameters are then used on the remainder of
%    the video: the test set. The desired false positive and false negative rates
%    are plotted against the realized rates on the test set.}
%  \label{fig:optimizer-targets}
%\end{figure}

\subsubsection{Running Time of the CBO}
\label{sec:eval-train-time}

\begin{figure}[t!]
  \centering
  \includegraphics[width=0.99\columnwidth]{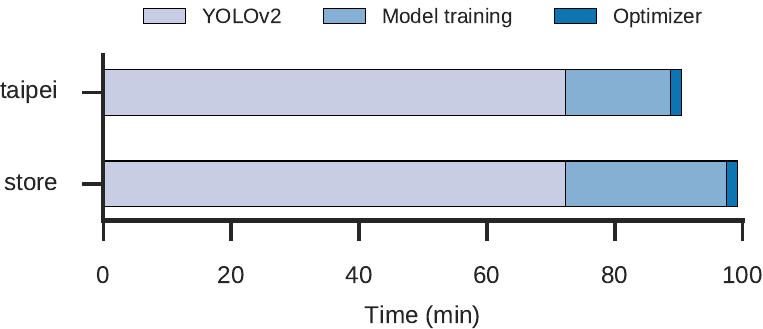}
  \vspace{1.0em}
  \caption{
  Breakdown of training and optimization time on \texttt{taipei} dataset.
  Passing all the training frames through YOLOv2 to obtain their true labels
  dominates the cost, followed by training all variants of our specialized
  models and then the rest of the steps in the CBO.
  }
  \vspace{-1em}
  \label{fig:train-time}
\end{figure}

We measured the time it takes to run our CBO across several
datasets, showing the most time-consuming one in
Figure~\ref{fig:train-time}.  In all cases, initializing \sn requires
labeling all the frames in the training data with YOLOv2, followed by
training all supported specialized models and difference detectors on
this data, then selecting a combination of them using the algorithm in
Section~\ref{sec:optimizer}. \colora{The CBO is efficient in the
  number of samples required: only 250k samples are required to train
  the individual filters and set the thresholds. For the longer
  videos, we randomly sample from the training set and for the shorter
  videos we use the first 250k frames.}  As shown in the figure,
YOLOv2 application takes longer than all the other steps combined,
meaning that \sn's CBO could run in real time on a second GPU
while the system is first observing a new stream.  Training of the
specialized NNs takes the next longest amount of time; in this case,
we trained 24 different model architectures.  We have not yet optimized this step or
tried to reduce the search space of models, so it may be possible
to improve it.

\subsection{Impact of Individual Models}
\label{sec:filter_eval}
\begin{figure}[t!]
  \centering
  \begin{subfigure}[b]{\columnwidth}
    \centering
    \includegraphics[width=\columnwidth]{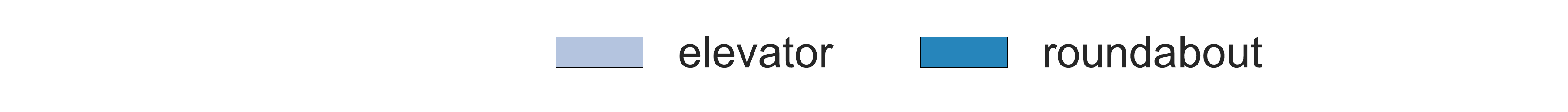}
    \vspace{-1.4em}
  \end{subfigure}
  \begin{subfigure}[b]{0.49\columnwidth}
    \centering
    \includegraphics[width=0.97\columnwidth]{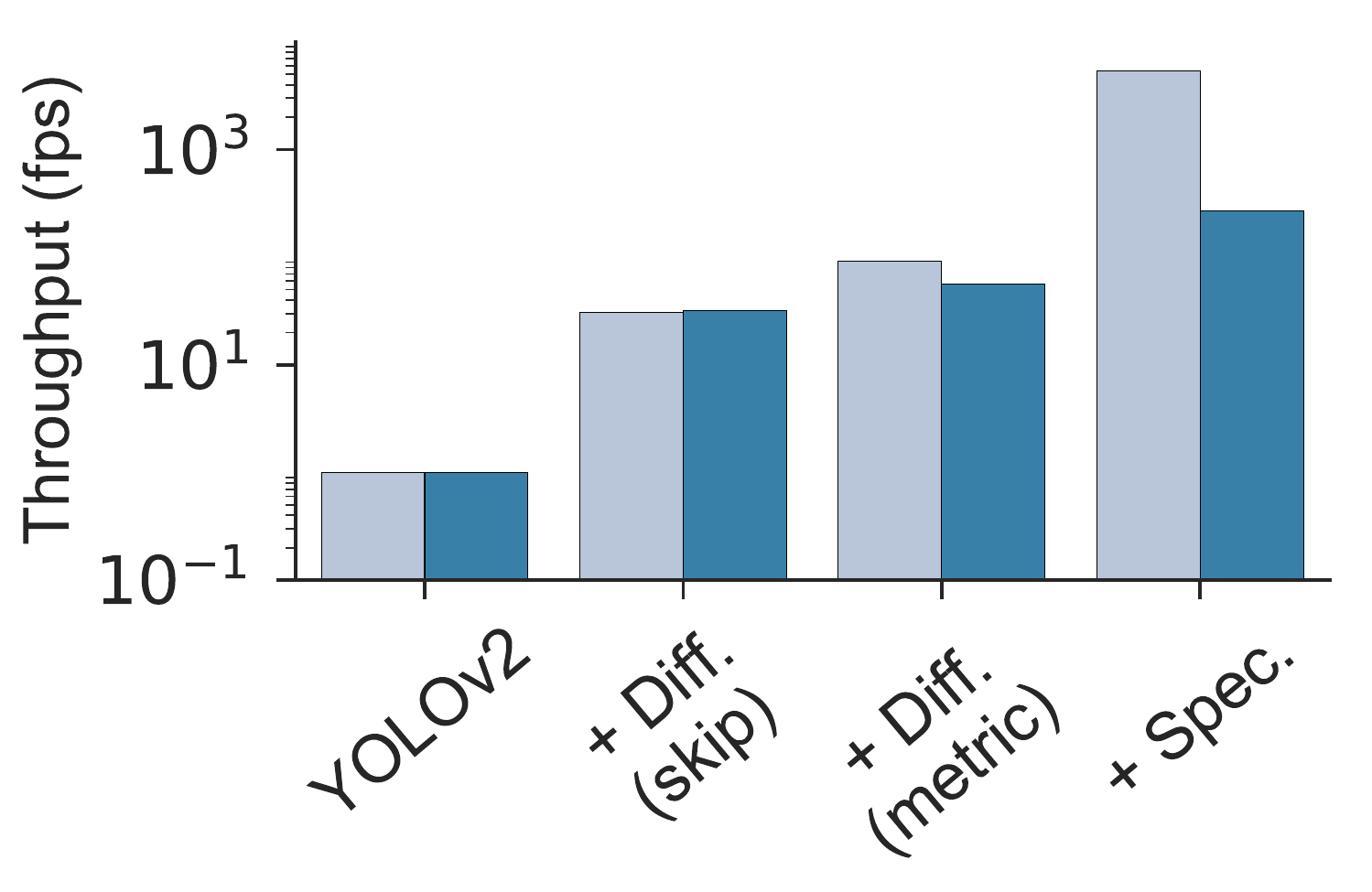}
    \vspace{-0.3em}
    \caption{Factor analysis}
    \label{fig:factor-analysis}
  \end{subfigure}
  \begin{subfigure}[b]{0.49\columnwidth}
    \centering
    \includegraphics[width=0.97\columnwidth]{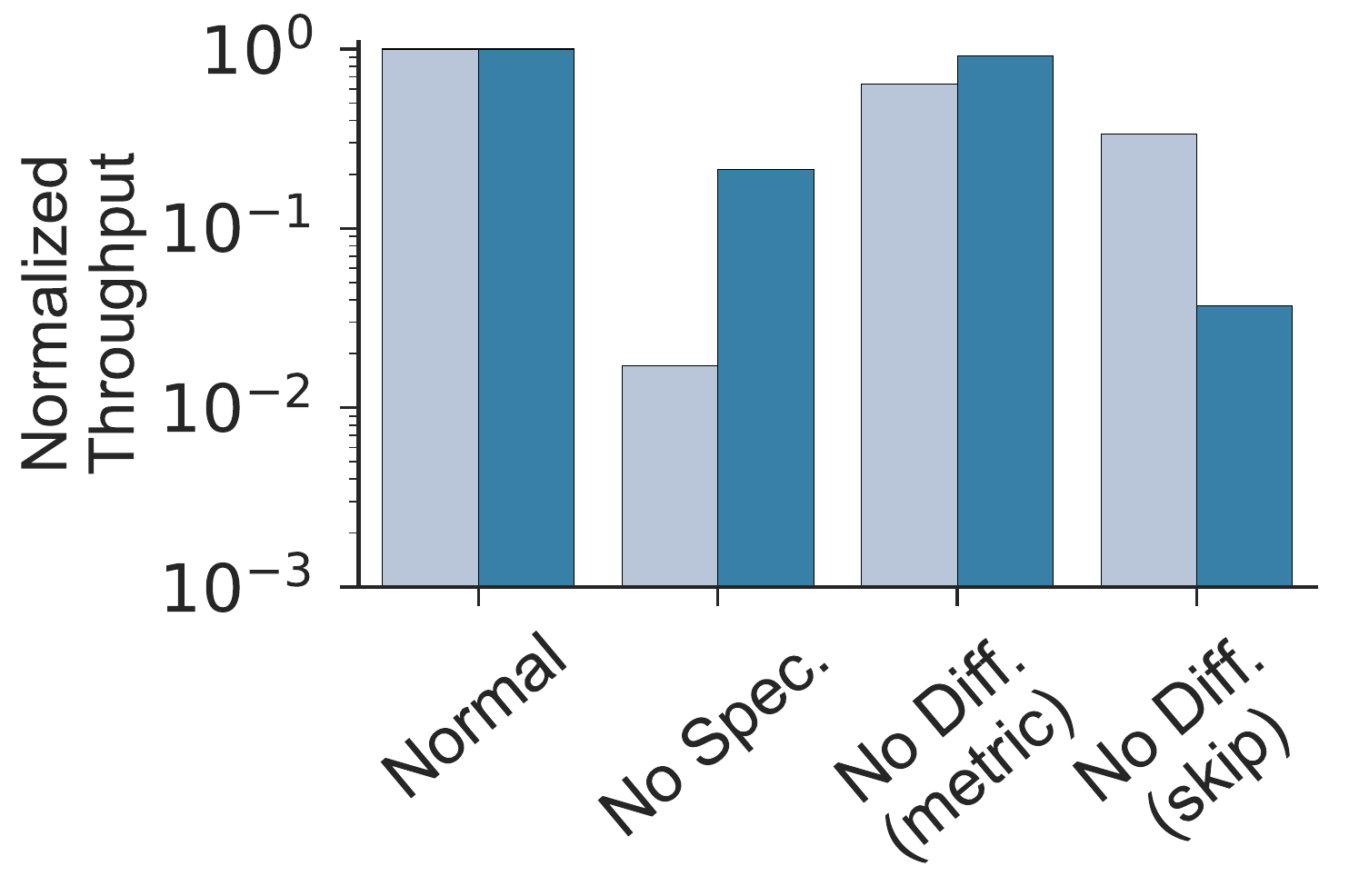}
    \vspace{-0.3em}
    \caption{Lesion study}
    \label{fig:lesion-study}
  \end{subfigure}\vspace{.5em}
  \caption{
  Factor analysis and lesion study of \sn's filters.  The factor analysis shows
  the impact of adding different filters for two videos; from left to right,
  we add each of the filters in turn over YOLOv2. The lesion study shows the
  impact of removing filters; the leftmost bars show normalized performance
  with all of \sn's features enabled, and the remaining bars to the right show
  the effect of removing each filter from \sn. (Note the logarithmic scale on
  the y-axes of both plots.)}
  \vspace{-0.5em}
\end{figure}

To analyze the impact of each of our model types on \sn's performance, we
ran a factor analysis and lesion study on two videos, with results shown
in Figures~\ref{fig:factor-analysis} and~\ref{fig:lesion-study}.

In the factor analysis, we started by running all frames through
YOLOv2 and gradually added: difference detection's frame skipping,
difference detection on the skipped frames, and specialized model
evaluation. Each filter adds a nontrivial speedup:
skipping contributes up to 30$\times$, content-based difference detection
contributes up to 3$\times$, and specialized models contribute up to
340$\times$.

In the lesion study, we remove one element at a time
from the complete \sn cascade.
As shown in Figure~\ref{fig:lesion-study}, each element contributes
to the overall throughput of the pipeline, showing that each component of \sn's cascades are important to its performance.

\subsection{Impact of Model Specialization}
\label{sec:eval-model-spec}
\begin{figure}[t!]
  \centering
  \includegraphics[width=0.99\columnwidth]{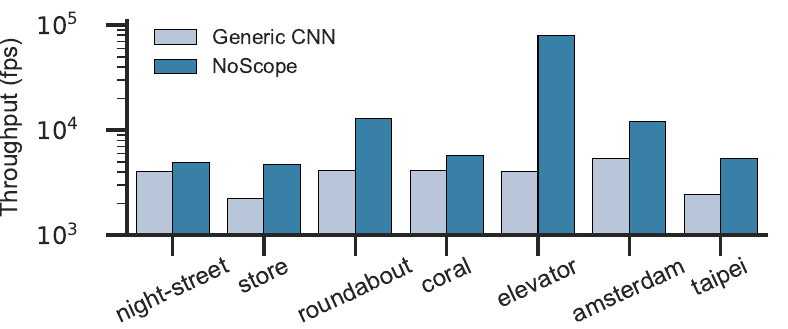}
  \vspace{1.0em}
  \caption{Throughput, Generic NN vs. \sn. Substituting the specialized NN
  model in \sn with an equivalent model trained on MS-COCO (a general-purpose
  training set of images used by YOLOv2) results in a decrease in the
  end-to-end throughput of the system across all videos.}
  \vspace{-1.0em}
  \label{fig:coco-sucks}
\end{figure}

Finally, we evaluate the benefit of video-specific model
specialization compared to training on general computer vision
datasets.  Our hypothesis in designing \sn was that we can achieve
much higher accuracy by training models on past frames from the
\emph{same video} to leverage the characteristics of that particular
scene (e.g., fixed perspective on the target object, fixed background,
etc.). To evaluate this hypothesis, we trained three deep NNs for
binary classification on the classes of objects we evaluate \sn on:
people, buses, and cars using the more general MS-COCO
dataset~\cite{ms-coco}, a recent high-quality object detection
dataset. For each class, we selected the best model from the same
model family as \sn's CBO.  Figure~\ref{fig:coco-sucks} shows
the resulting throughput across our videos.  In all cases, the
specialized models trained by \sn outperform the generic model of the
same size trained on MS-COCO (up to 20$\times$), showing that scene-specific
specialization has a  significant impact on designing models for efficient inference.

\subsection{\colora{Comparison Against Baselines}}
\label{sec:eval-baselines}
\begin{figure}[t!]
  \centering
  \includegraphics[width=0.99\columnwidth]{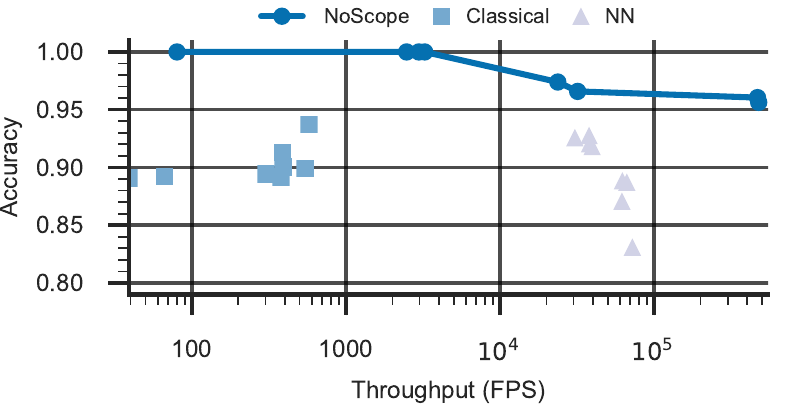}
  \vspace{0.5em}
  \caption{
  Comparison against classical and NN baselines. Experiments are run on \texttt{elevator} with frame skipping enabled for all models.
  }
  \vspace{-0.5em}
  \label{fig:baselines}
\end{figure}

\colora{We also compared to classic methods in computer vision,
  including a deformable parts model (which performed favorably in the
  ImageNet 2012~\cite{russakovsky2015imagenet}), a variety of
  SIFT-based bag-of-words classifiers (varying the SIFT parameters),
  and selective search with a SIFT-based classifier \cite{sift} (which performed
  well on PASCAL VOC~\cite{pascal-voc-2012}). We use implementations
  from OpenCV, a reference standard optimized computer vision library.
  Additionally, we trained several classification deep networks on the
  MS-COCO~\cite{ms-coco} dataset, varying hyper-parameters (2-10
  convolutional layers and 1-2 fully connected
  layers). Figure~\ref{fig:baselines} illustrates the resulting
  accuracy-speedup trade-offs for these models combined with frame
  skipping. \sn is nearly a thousand times faster than several classical
  baselines while achieving higher accuracy. Additionally,
  non-specialized deep networks also fail to achieve comparable
  accuracy. Thus, specialization is key to accurate and
  efficient inference, and \sn provides a means of smoothly trading accuracy
  for speed.}

%!TEX root=vuse.tex
\section{Related Work}
\label{sec:relatedwork}

\sn builds on a long tradition of data management for multimedia and
video and upon recent advances in computer vision, machine learning,
and computer architecture. We outline several major related thrusts
below.

\minihead{Visual Data Management} Data management researchers have
long recognized the potential of computer vision techniques for
organizing and querying visual data---starting with early systems such
as Chabot~\cite{chabot} and QBIC~\cite{qbic}, and followed by a range
of ``multimedia'' databases for storing~\cite{video-db1,vdb-index},
querying~\cite{video-mining,video-db2,aref-vid,jacob-db}, and
managing~\cite{av-db,video-db3,mm-survey} video data.  Similar
techniques are deployed in commercial software such as Oracle
Multimedia and Google Image Search. Many of these systems use classic
computer vision techniques such as \emph{low-level} image features
(e.g., colors, textures), and many of them rely on
augmenting video with text for richer semantic queries; however, with
the advent of neural networks, we believe it is worthwhile to revisit
many of these system architectures.

\minihead{Computer Vision Tasks} In the design of its CBO and model
cascade, NoScope draws on a range of computer vision techniques:

\miniheadit{Model Cascades} \colora{The concept of combining a
  sequence of classifiers to improve inference speed is known as a
  \emph{cascade}. One of the first cascades, the Viola-Jones
  detector~\cite{viola-jones}, cascades traditional image processing
  features; if any classifier is confident about the output, the
  Viola-Jones cascade short-circuits evaluation, improving execution
  speed.  Recent work has focused on learning cascades for pedestrian
  detection~\cite{learning-cascades} and facial
  recognition~\cite{cascade-face,cascade-face2} \emph{image} tasks. In
  \sn, we design a cascade architecture for use in \emph{video},
  specifically adapted to account for temporal locality (via
  difference detectors) and query simplicity (via
  model specialization). Similar to \sn, several NNs specialized for
  video take multiple frames as input to improve
  accuracy~\cite{cnn-multi,cnn-multi-2}, but again, these NNs focus on
  accuracy, not inference speed.  Finally, \sn's use of a cost-based
  optimizer to optimize the end-to-end cascade (i.e., selecting
  different filters and choosing different thresholds for each filter
  based on the video stream and a reference NN) is novel compared to
  work in the computer vision literature that uses the same cascade of
  filters (or that optimize solely for accuracy, not computational
  cost; cf.~\cite{boosted}). In this way, \sn acts as a system for
  model search~\cite{modelsel1, modelsel2,zoph2016neural} for
  efficient inference and is inspired by conventional database cost
  optimization~\cite{systemr} and related
  efforts on self-adapting query processing
  engines~\cite{cracking,selftuning}.}

\miniheadit{\colora{Video Object Detection/Extraction and Tracking}}
\colora{The overall task of object detection and extraction is a core
  task in computer vision. As we have discussed, the explosion of
  recent interest in deep networks for this task have largely focused
  on improving accuracy. Object detection in video is becoming
  increasingly popular: since 2015, the ImageNet competition has had a
  video detection task~\cite{russakovsky2015imagenet}.
  State-of-the-art methods~\cite{kang2016object, kang2016tcnn,
    seqnms2016} follow a common pattern: run still image detection
  over every frame of the video and post-process the result.  As still
  image detection is a building block of these methods, they are not
  optimized to exploit forms of temporal locality and instead
  primarily focus on accuracy; in contrast, \sn focuses on the
  trade-off between accuracy and inference speed.

  Beyond object detection, object tracking refers to the task of
  tracking an object through a video and is a vital tool in video
  analysis~\cite{yilmaz2006object}. There are many such methods for
  tracking, including NN-based methods~\cite{nam2016modeling, chi2017dual}
  and traditional methods~\cite{danelljan2016beyond, du2016online},
  that all performed well on the VOT2016 challenge~\cite{kristan2016}.
  In a recent survey of
  object tracking methods, the fastest method runs at only 3.2k
  fps~\cite{wu2013online}, and the above methods run much slower.
  \sn solves the simpler task of binary
  classification, and we see extending \sn to detection as a valuable
  area for future work.}

\miniheadit{\colora{Video Monitoring}} \colora{Video monitoring
  captures a variety of tasks~\cite{kastrinaki2003survey}, ranging
  from object detection~\cite{kim2003efficient} to vehicle
  tracking~\cite{tian2011video}. Historically, these systems have been
  tailored to a specific task (e.g. license plate
  detection~\cite{anagnostopoulos2008license},
  counting cars~\cite{zhang2017live},
  tracking pedestrians or cars~\cite{babenko2011robust}).
  Our goal in this work is to provide an automatic and
  generic framework for tailoring reference NNs to new tasks.}

\miniheadit{\colora{Scene Change Detection}} \colora{\sn leverages
  difference detectors to determine when labels for video contents
  have likely changed. This task is closely related to the task
  of scene change detection, in which computer vision techniques
  highlight substantial changes in video (e.g., cuts or fades in
  movies)~\cite{jiang1998scene, brunelli1999survey}. \sn's use of
  difference detection is inspired by these techniques but differs in
  two critical respects. First, \sn's difference detectors highlight
  changes in \emph{label} (e.g., bus enters), not just changes in
  scene. Second, \sn works over fixed-angle cameras. We expect many of
  these prior techniques to be useful in adapting \sn to moving
  cameras.}

\minihead{\colora{Image Retrieval}} \colora{Image retrieval is closely
  related to image classification. Retrieval typically takes one of
  two forms. The first is to associate images with a specific
  object or place of interest (e.g., this image is a photograph of a
  specific building on Stanford's
  campus~\cite{philbin2007object}). This first class of techniques are
  not directly applicable to our target setting: for example, in the
  night-street query, our task is to identify the presence of
  \textit{any} car, not a specific car. The second class of techniques
  allows retrieval of similar images from a large corpus that are
  similar to a query image~\cite{lin2015deep, yue2015exploiting}
  (e.g., Google Reverse Image search). One could use retrieval
  techniques such as k-nearest neighbors to perform binary
  classification. However, in this work, we leverage state of the art
  models---i.e., deep neural networks---that are specifically trained
  for the binary classification task.}

\minihead{Improving Deep Network Speed} Two recent directions in
related work study the problem of accelerating deep network inference
time:

\miniheadit{Model Compression and Distillation} Model
compression/distillation, i.e. producing a smaller, equivalent
model from a larger model, has been proposed to reduce the size and
evaluation time of NNs.
One strategy for model
compression is to use the behavior of a larger model (e.g., its output
confidences) to train a smaller model, either distilling an ensemble
of models into a smaller model~\cite{caruana-compression}, or
distilling one large network into one smaller
network~\cite{jeff-distill,ba-distill}. 
Another strategy is to modify
the network weights directly, either via
hashing~\cite{hashing-compress} or pruning and literal compression
of the weights
themselves~\cite{han-compression,xor-net}. These networks are almost
always more amenable to hardware acceleration---a related line of
research~\cite{han-eie,minerva}. The key difference between this
related work and our model specialization method is that this related
work is designed to produce smaller models preserve the full
\emph{generality} of the large model: that is, the small model can perform
all of the tasks as the large model, without any loss in accuracy.
In contrast, \sn explicitly \emph{specializes} NNs that were trained
on a large variety of input images and target objects to instead
operate over a specific video with a smaller number of target
objects; this specialization is key to improving
performance.

\miniheadit{Model Specialization} We are not aware of any other work
accelerating NN inference time via task specialization on
video. However, related strategies for specialization have been shown
to boost \emph{accuracy} (but not
runtime)~\cite{specialization-1, specialization-2}, and there are a
range of models designed for detecting specific classes of objects in
video. One of the most popular video-based object detection tasks is
pedestrian detection~\cite{ped-survey,ped-2}, with a range of
NN-based approaches (e.g.,~\cite{deepnet-ped}). In \sn, our goal is
to specialize to a \emph{range} of inputs automatically, using a
large general NN as guidance. We evaluate a handful of different labels
based on available datasets, but theoretically any class
recognized by NNs today could be specialized.

\minihead{ML Model Management and Stream Processing Systems} While we
have provided an overview of related multimedia query engines
and a comparison with several filter- and
UDF-based query processors in Section~\ref{sec:optimizer}, \sn also
draws inspiration from a number of recent machine learning model
management systems. Velox~\cite{velox} provides a high-performance
platform for serving personalized models at scale, while
MLBase~\cite{mlbase} provides a declarative interface for specifying
modeling tasks, and MacroBase~\cite{mb-cidr,mb-overview} provides a
platform for executing classification and data aggregation tasks on
fast data. Similarly, a range of systems from academia and industry
provide stream processing
functionality~\cite{borealis,telegraphcq,gigascope,wavescope}; here,
we specifically focus on video stream processing.  Like
Bismarck~\cite{bismarck}, \sn is structured as a sequence of dataflow
operators, but for performing video classification.  Of these
systems, MacroBase is the most closely related; however, in \sn, we
focus specifically on video classification and develop new operators and
a CBO for that task. Looking forward, we are eager to integrate \sn's
classifiers into MacroBase and similar systems.

%!TEX root=vuse.tex
\section{Conclusions}
\label{sec:conclusion}

Video is one of the richest and most abundant sources of data
available. At the same time, neural networks (NNs) have dramatically
improved our ability to extract semantic information from video.
However, na\"ively applying NNs to detect objects in video is
prohibitively expensive at scale, currently requiring a dedicated GPU
to run at real-time.  In response, we presented \sn, a system for
accelerating inference over video at scale by up to three orders of
magnitude via \emph{inference-optimized model search}. \sn leverages two
types of models: specialized models that forego the generality of
standard NNs in exchange for much faster inference, and difference
detectors that identify temporal differences across frames. \sn
combines these model families in a cascade by performing efficient
cost-based optimization to select both model architectures (e.g.,
number of layers) and thresholds for each model, thus maximizing
throughput subject to a specified accuracy target. Our \sn prototype
demonstrates speedups of two- to three orders of magnitude for binary
classification on fixed-angle video streams, with 1-5\% loss in
accuracy. These results suggest that by prioritizing inference time in
model architecture search, state-of-the-art NNs can be applied to
large datasets with orders of magnitude lower computational cost.

\section*{Acknowledgements}
We thank the many members of the Stanford InfoLab for their valuable
feedback on this work. This research was supported in part by
affiliate members and other supporters of the Stanford DAWN
project---Intel, Microsoft, Teradata, and VMware---as well as
industrial gifts and support from Toyota Research Institute, Juniper
Networks, Visa, Keysight Technologies, Hitachi, Facebook, Northrop
Grumman, and NetApp and the NSF Graduate Research Fellowship under
grant DGE-1656518.

\scriptsize

\balance

\bibliographystyle{abbrv}
\bibliography{refs}  % sigproc.bib is the name of the Bibliography in this case

\end{document}